\documentclass[aps, longbibliography, twocolumn, nofootinbib]{revtex4-2}
\usepackage{amsmath} 
\usepackage{graphicx}
\usepackage{dcolumn}
\usepackage{bm}
\usepackage{caption}
\usepackage{subcaption}
\usepackage{comment} 
\usepackage[dvipsnames]{xcolor} 
\usepackage{hyperref}
\usepackage[nameinlink, english]{cleveref} 
\usepackage{enumerate}
\usepackage{amsmath}
\usepackage{amssymb}
\usepackage[f]{esvect}
\usepackage{comment}

\begin{document}

--------------------------------------------------------------------
\title{Geometric structure of parameter space in immiscible two-phase flow in porous media}

\author{Håkon Pedersen\email{hakon.pedersen@ntnu.no}} \author{Alex
  Hansen \email{alex.hansen@ntnu.no}} \affiliation{PoreLab, Department of
  Physics, Norwegian University of Science and Technology, NO--7491 Trondheim,
  Norway}
\date{\today {}}
\begin{abstract}
  In a recent paper, a continuum theory of immiscible and incompressible
  two-phase flow in porous media based on generalized thermodynamic principles was
  formulated (Transport in Porous Media, 125, 565 (2018)). In this theory, two
  immiscible and incompressible fluids flowing in a porous medium are treated as
  a single effective fluid, substituting the two interacting subsystems for a
  single system with an effective viscosity and pressure gradient. In assuming
  Euler homogeneity of the total volumetric flow rate and comparing the
  resulting first order partial differential equation to the total volumetric
  flow rate in the porous medium, one can introduce of a novel
  velocity that relates the two pairs of velocities. This velocity, the
  co-moving velocity, describes the mutual co-carrying of fluids due to
  immiscibility effects and interactions between the fluid clusters and the porous medium 
  itself. The theory is based upon general principles of
  classical thermodynamics, and allows for many relations and analogies to draw upon in
  analyzing two-phase flow systems in this framework. The goal of this work is
  to provide additional connections between geometric concepts and the
  variables appearing in the thermodynamics-like theory of two-phase flow. In this
  endeavor, we will encounter two interpretations of the velocities of the
  fluids: as tangent vectors (derivations) acting on functions, or as coordinates on
  an affine line. The two views are closely related, with the former viewpoint being
  more useful in relation to the underlying geometrical structure of equilibrium
  thermodynamics, and the latter being more useful in concrete computations and 
  finding examples of constitutive relations. We apply these relatively 
  straightforward geometric contexts to interpret the relations
  between velocities, and from this obtain a general form for the co-moving
  velocity. 
  \end{abstract}
\maketitle
\section{Introduction}
\label{sec:intro}

The formulation of an effective continuum-level theory of immiscible and
incompressible two-phase flow in porous media based on rigorous physical
principles is a problem of great importance spanning several disciplines within
physics and mathematics 
\cite{bearDynamicsFluidsPorous1988, sahimi2011flow,bluntMultiphaseFlowPermeable2017, federPhysicsFlowPorous2022}.
The behavior of such flows underpins a range of complex phenomena seen in nature, 
industrial applications and general theoretical models where one can map a problem onto a
description where two interacting populations, here being fluids, are
exploring a constrained and complex network.  

In the continuum limit of these complex systems, which we have to consider 
in order to understand diverse phenomena such as permeability, wettability properties, 
capillary pressure and many more \cite{bluntMultiphaseFlowPermeable2017}, the macroscopic behavior
depends on the interactions at the pore scale, which in turn depend on
those at even smaller scales. Moreover, upon
coarse-graining of a system, new behavior might emerge as a result of the interactions between
the coarse-grained constituents that make up the system
\cite{andersonMoreDifferent1972}. The choice of how one ``abstracts away'' 
the behavior of a system at some scale to yield a new system at a larger length scale 
depends on the available information about the physical system.

In the continuum limit where it no longer makes sense to regard the system as
an immiscible mixture of two fluids moving in a solid matrix, but rather as 
a single fluid with complex rheological properties, one is interested in what the
behavior of this system is as a function of a minimal set of macroscopic
variables given that we know the individual properties of the component fluids
and the porous medium. This is the same question which is asked in
equilibrium thermodynamics
\cite{callenThermodynamicsIntroductionThermostatistics1985}, where the
equilibrium state is assumed to be uniquely defined by a set of macroscopic
variables, but reflecting the underlying molecular system through an equation of state. 

The notion of a state is important here.  In the context of immiscible two-phase
flow in porous media, its meaning is that the flow is determined by the current values of
the macroscopic variables alone without being dependent on how the system got there
\cite{royCoMovingVelocityImmiscible2022}.  This has been investigated experimentally and
computationally \cite{erpelding2013history}, leading to the conclusion that indeed 
steady-state flow is uniquely described by the values of the macroscopic variables.
However, there are regions where there is history dependence in the form of hysteresis
\cite{holtzman2020origin}. This means that there are regions in the space spanned by
the control variables where flow is described by functions that are multivalued
\cite{poston2014catastrophe}. This is
analogous to thermodynamics where hysteresis e.g., is a typical feature of first
order phase transitions. 

Two-phase flow in porous media is dissipative on the molecular scale, so it is
not in equilibrium in the sense of molecular thermodynamics. However,
we are free to define another notion of equilibrium based on properties of the
overall flow, for instance based on volumetric flow rates.  In doing so, we 
may map steady-state flow onto an equivalent equilibrium system 
\cite{hansenStatisticalMechanicsFramework2023}.
Under steady-state conditions, the fluid flow should then be determined by a set of thermodynamics-like
variables. With this general approach, one can map theories onto the general
framework of thermodynamics and leverage thermodynamic identities to obtain
relations between variables. 

This line of reasoning has been pursued in a series of papers that use statistical mechanics 
to scale up the immiscible two-phase flow problem from the pore scale to the continuum
scale, leading to a thermodynamics-like mathematical structure at the continuum level
\cite{hansenRelationsSeepageVelocities2018,royFlowAreaRelationsImmiscible2020,
  royCoMovingVelocityImmiscible2022, hansenStatisticalMechanicsFramework2023}.
As a consequence of this approach, immiscible two-phase flow in porous media
provides sensible physical analogues of quantities like energy, entropy, temperature etc.\ 
and relations like the Maxwell relations and the framework of phase transitions --- however in
a non-thermal setting.
  
Thermodynamics is at its core constrained by relations which manifest themselves in 
{\it geometric terms.\/} As long as one has a notion of equilibrium and seeks a minimal set 
of variables, the problem may be formulated as a problem of geometry 
\cite{weinholdMetricGeometryEquilibrium1975a,
ruppeinerRiemannianGeometryThermodynamic1995,
andresenThermodynamicGeometryMetrics1988, hermannGeometryPhysicsSystems1973}. The central
idea of the present paper is to use the analogue between immiscible two-phase flow in 
porous media and thermodynamics to use the geometrical apparatus previously implemented for 
thermodynamics in the context of the immiscible two-phase flow problem.  

\subsection{Immiscible two-phase flow in porous media formulated as a thermodynamic problem}
\label{sec:pseudo-therm-intro}

In Hansen et al.\ \cite{hansenRelationsSeepageVelocities2018}, Euler homogeneity was used to
formulate thermodynamic relations for the steady-state seepage velocities $v_w$ and 
$v_n$ of two immiscible and incompressible fluids (respectively more wetting and less wetting 
with respect to the solid matrix).  Central to this work was to provide a mapping between 
the two seepage velocities and the average seepage velocity $v$.  By averaging, we would map
$(v_w,v_n)\to v$, but it is not possible to construct a unique inverse mapping, $v \to (v_w,v_n)$.
This led to the introduction of the {\it co-moving velocity\/} $v_m$ so that 
$(v,v_m) \to (v_w,v_n)$, the inverse mapping.  The two mappings are
\begin{eqnarray}
v&=&S_w v_w+S_n v_n\;,\label{eq-ah-1}\\
v_m&=&S_w\frac{\partial v_w}{\partial S_w}+S_n\frac{\partial v_n}{\partial S_w}\;,
\label{eq:co-moving-convex-comb}
\end{eqnarray}
and
\begin{eqnarray}
v_w&=&v+S_n\left[\frac{\partial v}{\partial S_w}-v_m\right]\;,\label{eq-ah-3}\\
v_n&=&v-S_w\left[\frac{\partial v}{\partial S_w}-v_m\right]\;,\label{eq-ah-4}
\end{eqnarray}
respectively. Here $S_w$ and $S_n$ are the wetting and non-wetting saturations respectively, obeying
\begin{equation}
\label{e10}
S_w+S_n=1\;.
\end{equation}
We will return to these equations in Section \ref{sec:rea}.

Why would one want to construct this inverse mapping, $(v,v_m) \to (v_w,v_n)$?  It was observed 
experimentally in 2009 \cite{tallakstad2009steady,tallakstad2009steadyb} that the average seepage 
velocity $v$ follows a power-law in the pressure gradient with an exponent considerably larger than 
one (as would be the case for Darcian flow) over a wide range of capillary numbers. This observation 
has been followed up in multiple papers since, see e.g., 
\cite{sinha2012effective,aursjo2014film,gao2017x,sinha2017effective,
gao2020pore,zhang2021quantification,zhang2022nonlinear,fyhn2021rheology,anastasiou2024steady}.
Experimentally, one finds this 
power-law behavior around a capillary number of the order of $10^{-5}$ and up. 
The power law appears when an increase in pressure gradient results in the mobilization of 
interfaces that would otherwise be held in place by the capillary forces. If we assume that
the increase in mobilized interfaces is proportional to the increase in pressure gradient
and the increase in effective permeability is proportional to the increase in mobilized
interfaces, we end up with an exponent equal to two. The flow rate-pressure gradient reverts to
being linear again when all interfaces that may move are moving \cite{sinha2017effective}. 
Having the mapping from $(v,v_m)$ to
$(v_w,v_n)$, equations (\ref{eq-ah-3}) and (\ref{eq-ah-4}), makes it possible to reconstruct the 
seepage velocity constitutive equations for each fluid from the constitutive equation between $v$ 
and the pressure gradient. 

This brings us to the co-moving velocity $v_m$, see equation (\ref{eq:co-moving-convex-comb}).  
We define the wetting saturation $S_w$
in the following way: We imagine a cut through the porous medium. Part of the
cut will go through the matrix and part will go through the pores. The area of the plane
cutting through the pores is the pore area $A_p$. The wetting 
saturation $S_w$ in that plane is the fraction of the pore area $A_p$ that cut through the
wetting fluid.  It was shown in \cite{hansenStatisticalMechanicsFramework2023} that a 
natural variable describing the co-moving velocity is the {\it flow derivative\/} 
$\mu=v^{\prime}=dv/dS_w$.
Both numerical and experimental data point towards the constitutive equation for
$v_m$ being quite simple \cite{federPhysicsFlowPorous2022,royCoMovingVelocityImmiscible2022,alzubaidi2023impact} and
Hansen has proposed that the origin of this simplicity may be found in dimensional
analysis \cite{hansen2024linearity}.  The constitutive equation seems to be an 
affine function of the form
\begin{equation}
  \label{eq:vm-constitutive}
  v_m \ = \ b v' + a v_0 \ \;,
\end{equation}
to within the accuracy of the measurements.  Here $a$ and $b$ are dependent on the viscosity ratio and pressure gradient \cite{royCoMovingVelocityImmiscible2022,hansenStatisticalMechanicsFramework2023}, and $v_0$ is a velocity scale. 

We will in this paper  in the context of geometry examine the two-way mappings 
\begin{align}
  (v,v_m) \ \leftrightarrow& \ (v_w,v_n) \ \label{eq:primary-transf} \\
  (\hat{v}_w,\hat{v}_n) \ \leftrightarrow& \ (v_w,v_n) \ \label{eq:secondary-transf} \;,
\end{align}
where the first mapping we have already described in equations (\ref{eq-ah-1}) to (\ref{eq-ah-4}).  
The second mapping (\ref{eq:secondary-transf}) is between the thermodynamic velocities, defined as
\begin{align}
  \hat{v}_{w}=&\left( \frac{\partial Q}{\partial A_{w}}\right)_{A_{n}} \;,
                \label{eq:vw-therm} \\
  \hat{v}_{n}=&\left( \frac{\partial Q}{\partial A_{n}}\right)_{A_{w}}  \;,  
                \label{eq:vn-therm}
\end{align}
and the seepage velocities. We have not explicitly written out the dependence on the pressure
gradient in these two expressions. Here 
$Q$ is the volumetric flow rate through the cut
described above, $A_w$ is the area of the cut passing through the wetting fluid and $A_n$ is the
area of the cut passing through the non-wetting area.  The thermodynamic velocities 
appear naturally in the thermodynamics-like formalism proposed in \cite{hansenRelationsSeepageVelocities2018,hansenStatisticalMechanicsFramework2023}.

Preliminary steps towards such a geometric interpretation was taken in
\cite{pedersenParameterizationsImmiscibleTwophase2023a}, where different
coordinate systems on the space spanned by the first quadrant of $(A_w,A_n)$
were defined and studied. This description was completely linear, in
the sense that all quantities had an interpretation as components of
vectors with respect to some coordinate system. These vectors are elements of
the tangent space to the space of extensive variables, and simple
relations in the dual space of cotangent vectors were also considered.

The relation to equilibrium thermodynamics enters through the
steady-state flow condition, which is the situation when the
macroscopic variables of the system fluctuate around well-defined average
values.  By a transformation we will discuss in Section \ref{sec:rea},
the flow problem, which is characterized by the production of molecular entropy,
can be mapped onto an equivalent equilibrium system by noting that the information entropy 
associated with the pore-level fluid flow configurations is not being produced
\cite{hansenStatisticalMechanicsFramework2023}. The maximum entropy principle may
then be used to formulate a statistical mechanics \cite{jaynesInformationTheoryStatistical1957}
which in turn leads to an equilibrium thermodynamics-like formalism at the continuum scale. 

Our goal is to formulate the two-phase-flow problem in a
manner suitable for geometric generalizations (see below). The reason for seeking this
connection is to impart validity to the claim that immiscible and
incompressible two-phase flow is readily describable using principles of
thermodynamics. This ties into the general problem of applying thermodynamic theories 
to mesoscopic systems as a whole \cite{regueraMesoscopicDynamicsThermodynamic2005,
grmelaContactGeometryMesoscopic2014}. Our reasoning is that if we are able to
embed the thermodynamics-like theory into the geometric framework of thermodynamics in
a satisfactory manner, we can use geometric tools to obtain thermodynamic-like
relations and possibly investigate thermodynamic-like processes in this system.

In the geometric interpretation of thermodynamics, the usage of mathematical
structures called fiber bundles
\cite{leeIntroductionSmoothManifolds2012, milnorCharacteristicClasses1974} are
prevalent. In this article, we will not go into much detail on these structures,
and only consider relatively common types of fiber bundles, in
particular the tangent vector bundle. While we will introduce these objects and remark where 
they are applicable, the structure itself will not be the main subject of this work.

More explicitly, when we speak of a ``geometric formulation of thermodynamics'',
we refer to the {\it contact geometric formulation of thermodynamics\/}
\cite{bravettiContactGeometryThermodynamics2019,simoesContactGeometrySimple2020}. 
A symplectic formulation of thermodynamics is also possible by
introducing additional gauge variables
\cite{balianHamiltonianStructureThermodynamics2001}. The contact- and symplectic
formulations are closely related \cite{arnoldMathematicalMethodsClassical1978},
but have different purposes. Contact geometry has long been recognized as an
appropriate geometric setting for both equilibrium- and non-equilibrium
thermodynamics, and has close ties to information theory and statistical
mechanics \cite{bravettiContactGeometryThermodynamics2019}. We will not consider
contact geometry in detail in this work other than a short comment in
Section \ref{sec:contact-geometry}. 

\subsection{Motivation and outline}
\label{sec:motivation-goal-limitations}

The core idea of this work is to reframe the theory initially presented in
\cite{hansenRelationsSeepageVelocities2018} using the basic concepts from
two related geometric viewpoints. The first one is the basic
differential geometry and (tangent) bundle structure of the configuration space
of extensive variables. The second one is a classical geometric view of the
velocities as points in an affine space. We will only need basic concepts from
both; the difficulty here is not mathematical, but rather lies in the physical
interpretation of the results. The geometric relations are motivated by the
particular form of the equations to be presented in Section \ref{sec:rea}.

We note that essentially all concepts used in
this work are common tools in parts of mathematical physics. Our view
is that a thorough introduction to these ideas are needed when put in the
context of a pseudo-thermodynamic theory of two-phase flow in porous media, a
field where primarily other techniques have been applied. 

The ``classical'' geometric viewpoint interprets the
values of the functions corresponding to the velocities introduced in Section 
\ref{sec:pseudo-therm-intro} as points in an affine space. The formulation in terms of 
differential geometry describes the velocities in the theory as (tangent) vector fields. 
We will see how both views, which uses many of the same types of spaces but with different
objects defined on them, can aid in our understanding of what the co-moving
velocity, equation (\ref{eq:co-moving-convex-comb}), represents, and how to potentially work with it. 
Moreover, we will see how this theory relates to a constitutive relation for the 
co-moving velocity 
\cite{royCoMovingVelocityImmiscible2022,alzubaidi2023impact,hansen2024linearity}.

The tangent-vector formulation can be seen in relation to previous works
\cite{pedersenParameterizationsImmiscibleTwophase2023a}. The difference here is
that the tangent vectors are considered as derivative operators, where the
action of the tangent vector fields on functions defined on the space
yields the velocities.

The structure of the article is as follows: in Section \ref{sec:rea}, we
present the preliminaries of the pseudo-thermodynamic theory of two-phase flow
\cite{hansenRelationsSeepageVelocities2018}. In Section \ref{sec:spaces}, we 
introduce the machinery of manifolds, tangent- and affine spaces, and bundles
constructed from these spaces. These bundles
are the natural habitats of the vector fields presented in this work. We will
also present the preliminaries of using affine spaces in the classical-geometric
viewpoint. In Section
\ref{sec:co-moving-affine-map}, we show how the co-moving velocity appears
in the two geometric viewpoints presented above, and how it relates to the
interpretation of the equations in Section \ref{sec:rea}. This is the main
part of this work, with the goal of clearing up what the relations in Section
\ref{sec:rea} are seemingly stating in geometric terms, formulate them in terms
of geometry, and show how the co-moving velocity obtained in this way relates to
already known relations.

Before summing up our results in Section \ref{sec:conclusion}, we will in
Section \ref{sec:discussion-related} comment shortly on the usage areas of the
results of Section \ref{sec:co-moving-affine-map}. Moreover, we comment on two
related topics to the concepts introduced in this work, namely how the
results are related to contact geometry, and the notion of a connection
on a bundle.

\section{Preliminaries and the Euler Homogeneous Function Theorem}
\label{sec:rea}

\begin{figure}[tbh]
  \centering \includegraphics[width=\linewidth]{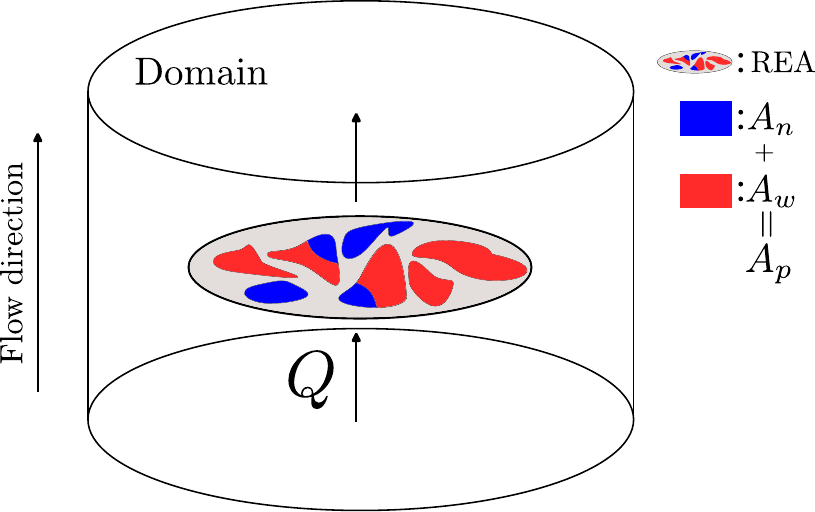}
  \caption[REA]{The porous medium sample with an REA indicated.  The pore area $A_p$
  can be divided into a wetting area $A_w$ and a non-wetting area $A_n$ so that their sum is 
  $A_p$.}
  \label{fig:REA}
\end{figure}

Consider a porous medium sample as shown in Figure \ref{fig:REA}.  We assume
the immiscible fluids enter through the bottom and leaving through the top.
The side walls are impenetrable. Within the porous medium, the fluids mix by forming clusters. The clusters merge and split, creating a steady-state.  We choose a plane orthogonal to the average flow direction far enough from the bottom so that it is in the region where the flow is in a steady-state.  In this plane we choose a Representative Elementary Area (REA)
which is large enough for the macroscopic variables to have well-defined averages, 
but not larger. The REA has an area $\tilde{A}$.  We use the tilde to signify that 
the area $\tilde{A}$ is the area of a single plane. Associated to the REA, there is 
a time averaged volumetric flow rate $Q$ of fluid passing through $\tilde{A}$ at each
instant. 

The average value of $\tilde{A}$ over the entire domain, defined as
the integral of $\tilde{A} = \tilde{A} (z)$, where $z$ is the coordinate along
the flow direction, is denoted by $A$. We will define all areas in this way, as
their averaged values over the domain in the overall direction
of $Q$ where the flow is in a steady state. We will in the following refer to the averaged area $A$ as the area of the REA.
We will in the following introduce several other kinds of areas.  These will in the same way be averages over sets of REAs. 

We define the porosity $\phi$ of the porous medium as
\begin{equation}
  \label{e4}
  \phi \ \equiv \ \frac{A_p}{A} \ \;,
\end{equation}
where $A_p$ is the area of $A$ that cuts through the pores. The solid matrix area $A_s$ is
given by $A_s= A \left(1-\phi \right)$. We assume the porous medium to be homogeneous.
The pore area $A_p$ is an
extensive variable; it scales with a factor $\lambda$ when we let $A \mapsto \lambda A$,
where $\lambda$ is a real number. The porosity $\phi$ does not change under under this
scaling, i.e., it is an intensive variable.

The pore area of the REA, $A_p$ is split into an area $A_w$ of (more) wetting fluid and an area $A_n$ of
(less) non-wetting fluid. The the fluids are taken to be incompressible. We have that
\begin{equation}
\label{e11}
 A_w+A_n= A_p\;.
\end{equation}
We then define the wetting and non-wetting saturations
\begin{eqnarray}
S_w=\frac{A_w}{A_p}\;,\label{eq-ah-10}\\
S_n=\frac{A_n}{A_p}\;,\label{eq-ah-11}
\end{eqnarray}
obeying equation (\ref{e10}). 

Since we consider the mutual flow of two fluids, $Q$ can be decomposed as a sum
of the volumetric flow rates of the individual fluids, denoted $Q_{w}$ and
$Q_{n}$. We then have 
\begin{equation}
  \label{eq:e6}
  Q(A_w, A_n)=Q_{w} \left( A_w, A_n\right)+Q_{n}\left( A_w, A_n\right)\;,
\end{equation}
so $Q$ may be seen as a composite thermodynamic-like system consisting of two subsystems. We define the seepage velocities as
\begin{align}
  v=&\frac{Q}{A_p} \label{e7}\;,\\
  v_w=&\frac{Q_w}{A_w} \label{eq:vw-seepage} \;,\\
  v_n=&\frac{Q_n}{A_n} \label{eq:vn-seepage}\;.
\end{align}
These velocities of the individual fluids passing through the REA are the ones measured in
experiments.

The {total} volumetric flow rate $Q$ is extensive in the variables $A_w$ and $A_n$, meaning that
\begin{equation}
  \label{eq:Q-homogeneity}
  Q(\lambda A_w,\lambda A_n)
  =\lambda Q(A_w,A_n)\;.
\end{equation}
We are here assuming $A_w$ and $A_n$ to be the control variables.  The pore area $A_p$ is then a dependent variable.  This is of course not possible to arrange in the laboratory.  However, theoretically it is possible. 

By defining $Q_w$, $Q_n$ in equation \eqref{eq:e6} as functions of $A_w$, $A_n$ and not as $Q_w(A_w)$ and 
$Q_n\left( A_n\right)$, we imply that $Q$ is not a sum of simple, non-interacting
subsystems \cite{callenThermodynamicsIntroductionThermostatistics1985}; the ``subsystem'' flow rates $Q_w$, $Q_n$ include interactions between
the two phases of fluids.  We could alternatively write $Q$ as the sum of two
non-interacting volumetric flow rates $Q_{w,0} $, $Q_{n,0}$ and an interaction term $Q_{\text{int}}$
\begin{equation}
  \label{eq:flow-rate-interaction-term}
  Q(A_w, A_n) \ = \ Q_{w,0}(A_w) +  Q_{n,0}(A_n) +
  Q_{\text{int}}\left( A_w, A_n\right) \;.
\end{equation}
However, we will keep to the convention in equation \eqref{eq:e6}, noting that this
assumption is a potential point of analysis in itself \cite{oppenheimThermodynamicsLongrangeInteractions}.

Using equations \cref{eq:e6} to \cref{eq:vn-seepage}, we find equation (\ref{eq-ah-1}). 
\Cref{eq:e6} can then be rewritten as
\begin{equation}
  \label{eq:Q-seepage}
  Q=A_wv_{w}+A_nv_{n}\;.
\end{equation}

\begin{figure}
  \includegraphics[width=\linewidth]{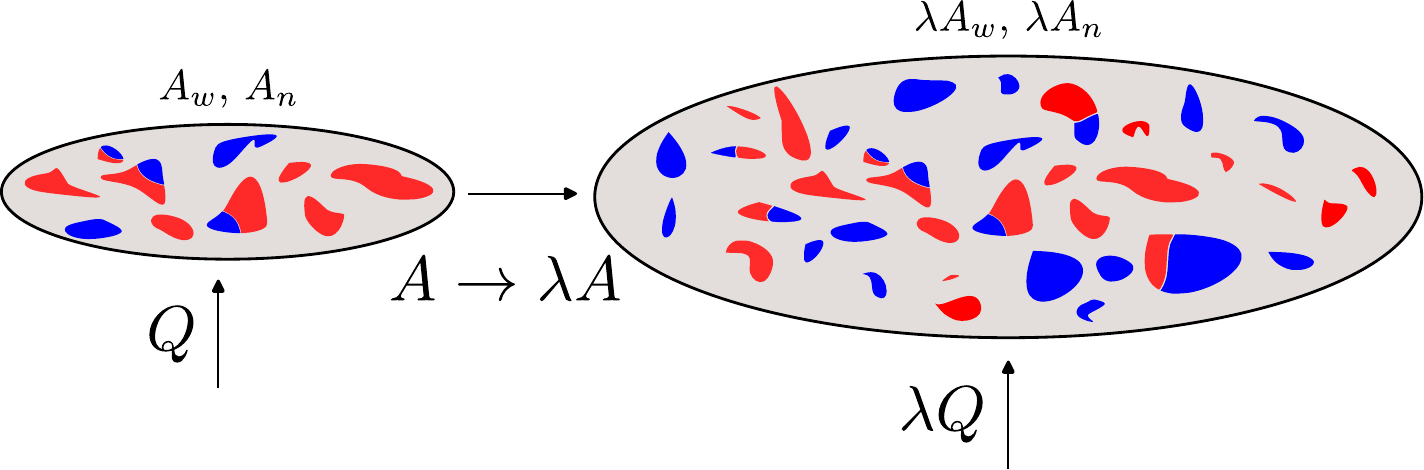}
  \caption{Scaling the area $A$ by a factor $\lambda$ scales the
    volumetric flow rate $Q$ in the same manner, demonstrating that $Q$ is an 
    Euler homogeneous function of degree one.}
  \label{fig:euler-scaling}
\end{figure}

We now use the assumption that $Q$ is degree one Euler homogeneous in the areas
\cite{hansenRelationsSeepageVelocities2018}. Taking the derivative with respect
to $\lambda$ on both sides of equation (\ref{eq:Q-homogeneity}) and setting $\lambda =1$, we get
\begin{equation}
  \label{eq:Q-euler-hom}
  Q(A_{w},A_{n})=A_{w}\left( \frac{\partial Q}{\partial A_{w}}\right)
  _{A_{n}}+A_{n}\left( \frac{\partial Q}{\partial A_{n}}\right)
  _{A_{w}}\;.
\end{equation}
By dividing equation (\ref{eq:Q-euler-hom}) by $A_{p}$, we get
\begin{equation}
  \label{eqn10-1}
  v=S_{w}\left( \frac{\partial Q}{\partial A_w}\right)
  _{A_n}+S_{n}\left( \frac{\partial Q}{\partial A_n}\right)_{A_w}\;.
\end{equation}
The partial derivatives acting on $Q$ have units of velocity, so we define the
{\it thermodynamic velocities\/} equations (\ref{eq:vw-therm}) and (\ref{eq:vn-therm}).
We may then write equation (\ref{eqn10-1}) as
\begin{equation}
  \label{eqn10-5}
  v=S_w \hat{v}_w+S_n\hat{v}_n\;.
\end{equation}
We will utilize the notation $ \hat{v}_i $ 
for the (set) $\left( \hat{v}_w, \hat{v}_{n} \right)$, and the same
(un-hatted) notation for the set of seepage velocities, $ v_i  \equiv
\left( v_w, v_n \right)$.

The thermodynamic velocities $\hat{v}_i$ are not the
same as the physical velocities $v_w$ and $v_n$. Rather, the most general
  relation   between $\left\{ \hat{v}_i \right\}$ and $\left\{ v_i \right\}$
that fulfills both equations (\ref{eq-ah-1}) and (\ref{eqn10-5}),
\begin{equation}
  \label{eq:v-both-definitions}
  v=S_w \hat{v}_w +S_n\hat{v}_n = S_wv_w+S_nv_n \ \;,
\end{equation}
is given by \cite{hansenRelationsSeepageVelocities2018}
\begin{align}
  \hat{v}_w \ =& \ v_w + S_n v_m \label{eq:vw-transf} \;, \\
  \hat{v}_n \ =& \ v_n - S_w v_m \label{eq:vn-transf} \;,
\end{align}
which defines the {\it co-moving velocity\/}, denoted $v_m$. Hence, the
co-moving velocity which first appeared in equation (\ref{eq:co-moving-convex-comb}) 
is a quantity with units of velocity that relates the thermodynamic
and seepage velocities.

It was shown in \cite{hansenRelationsSeepageVelocities2018} that
\begin{equation}
  \label{eq:vm-derivative-form}
  v_m + v_w - v_n  =  \hat{v}_w - \hat{v}_{n} 
  =  v^{\prime} \;,
\end{equation}
where $v^{\prime} = \mathrm{d} v / \mathrm{d} S_{w} $, which will be used throughout
this work.

One can show \cite{hansenRelationsSeepageVelocities2018} that $\hat{v}_i$ satisfies an analogue of the
{Gibbs-Duhem} relation,
\begin{equation}
  \label{eq:GD}
  S_w \left( \frac{\mathrm{d} \hat{v}_{w}}{\mathrm{d} S_w} \right) + S_n\left( \frac{\mathrm{d} \hat{v}_{n}}{\mathrm{d} S_w} \right) \ = \ 0 \ \;.
\end{equation}
The interpretation is, like in classical thermodynamics, that the intensive
thermodynamic velocities are fully dependent. In the same work, it was
shown that $v_m$ can also be expressed as equation (\ref{eq:co-moving-convex-comb}). Equations (\ref{eq-ah-1}) and
(\ref{eq:co-moving-convex-comb}) constitute the transformation $(v_w,v_n) \to
(v,v_m)$. From the above relations, one can show that 
\begin{align}
  \label{eq:partial-property-w}
  \hat{v}_w \ =& \  v + S_n \frac{d v}{d S_{w}} \;, \\
  \hat{v}_n \ =& \  v - S_w \frac{d v}{d S_{w}} \;. \label{eq:partial-property-n} \\
\end{align}
Combining these two equation with equations (\ref{eq:vw-transf}) and (\ref{eq:vn-transf}) leads to equations
(\ref{eq-ah-3}) and (\ref{eq-ah-4}), constituting the transformation $(v_p,v_m) \to
(v_w,v_n)$.

As already discussed, the constitutive equation for $v_m$ (\ref{eq:vm-constitutive}) 
is to within the precision of the measurements an affine function of $v'= dv/dS_w$.  

\section{Spaces and manifolds}
\label{sec:spaces}

We will in this section describe the theory presented in Section \ref{sec:rea} using 
manifolds and bundle structures.

\begin{figure}[h]
  \centering
  \includegraphics[width=\linewidth]{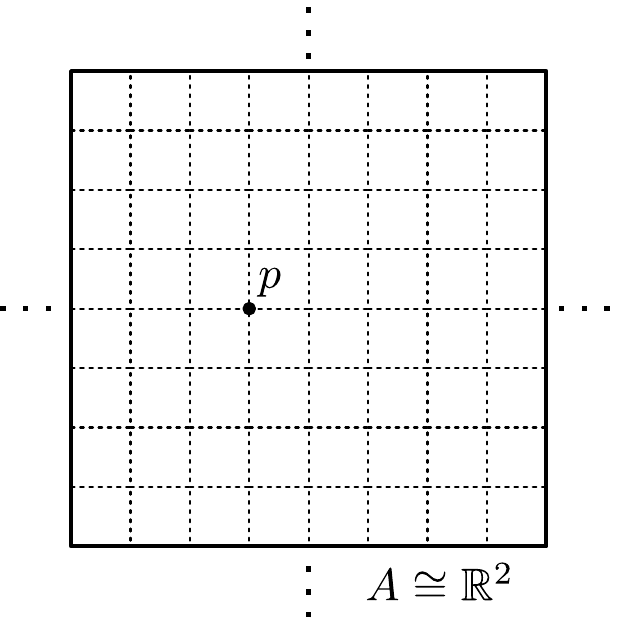}
  \caption{The manifold spanned by the extensive areas $A_w$, $A_n$ is taken to
    be an open subset of $\mathbb{R}^2$.}
  \label{fig:manifold-A}
\end{figure}

In \cite{pedersenParameterizationsImmiscibleTwophase2023a}, a 
two-dimensional vector space of the extensive area variables $(A_w,A_n)$ was
studied, and the terminology of manifolds was left out. The idea here is similar,
but we instead define the space of extensive areas to be a two-dimensional manifold 
where $(A_w, A_n)$ is a possible set of coordinates labeling a point on the 
manifold, see Figure \ref{fig:manifold-A}. We label this manifold by $\mathcal{M}$. 
Since we have from equation \eqref{e11} that $A_p$ is a dependent variable, we only
need two independent extensive variables as coordinates on $\mathcal{M}$. We choose 
them to be $A_w$ and $A_n$.

\begin{figure}
  \centering
 \includegraphics[width=\linewidth]{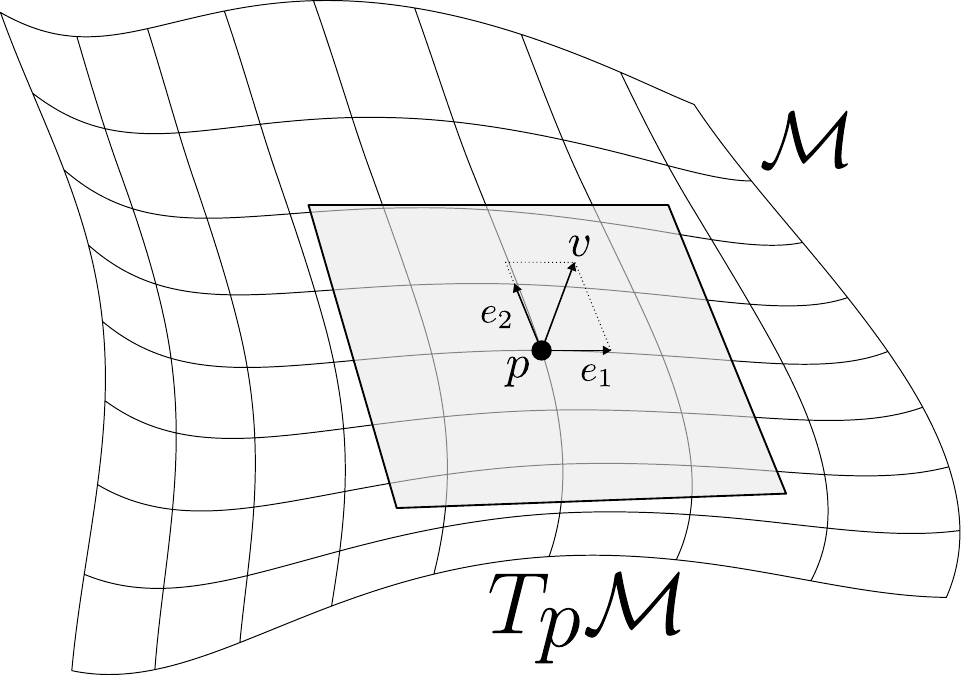} 
  \caption{The tangent space $T_p{\cal M}$ at a point $p \in {\cal M}$ can be imagined as a plane
    (strictly speaking a vector space)
    attached to $\cal M$ at the point $p$. A tangent vector $v \in T_p{\cal M}$ can be imagined 
    as a ``small arrow'' tangent to
    the manifold. The tangent vector $v$ can be expressed in some basis, for
    instance $\left( \mathbf{e}_1, \mathbf{e}_2 \right)$. }
  \label{fig:tangent-space}
\end{figure}

$\mathcal{M}$ itself does not have the structure of a vector space.  However, the tangent
space at each point of the manifold, which is just the space of all tangent vectors
that has this point as their initial point or origin, has such a structure, 
see Figure \ref{fig:tangent-space}. 
Since our space of extensive variables is essentially just $\mathbb{R}^2$, it might seem
unnecessary to separate the manifold from its tangent space. However, we
cannot come to any of the conclusions in this work if we do not formally keep
them separate.

The motivation for introducing a manifold and its tangent spaces is to be able
to formally discern extensive and intensive variables. This is
necessary to explain  why our theory acts like a thermodynamic theory. As mentioned
earlier, the vector spaces in
\cite{pedersenParameterizationsImmiscibleTwophase2023a} did not separate between
the space of extensive variables and that of velocities; areas and velocities
were simply elements of the same vector space. In a geometrical approach to
physics, one often separates the two by means of a bundle structure, with a
base-manifold acting as a configuration space, and some space of objects
attached to each point of the configuration space. 
The geometry of classical mechanics as a whole is based on this structure, 
and geometric descriptions of thermodynamics use
exactly the same framework. For instance, what we call ``extensive'' and
``intensive'' variables in thermodynamics are examples of canonical
coordinates \cite{arnoldMathematicalMethodsClassical1978}, the coordinates on
the ``thermodynamic phase space'' analogous to the phase space of positions and
momenta in Hamiltonian mechanics. Without a clear distinction between the two
types of variables, we will not be able to introduce geometric structures that
define thermodynamic equilibrium states, so called
Legendre-manifolds \cite{arnoldMathematicalMethodsClassical1978}, or
talk about metrics on the thermodynamic phase space, which connect 
thermodynamics to statistical mechanics \cite{mrugalaContactMetricStructures2000}.
Thus, separating the extensive and intensive variables in the same way as in
geometrical physics is a natural step in a ``geometrization'' of the theory in
this work.

\begin{figure}[h]
  \centering
  \includegraphics[width=\linewidth]{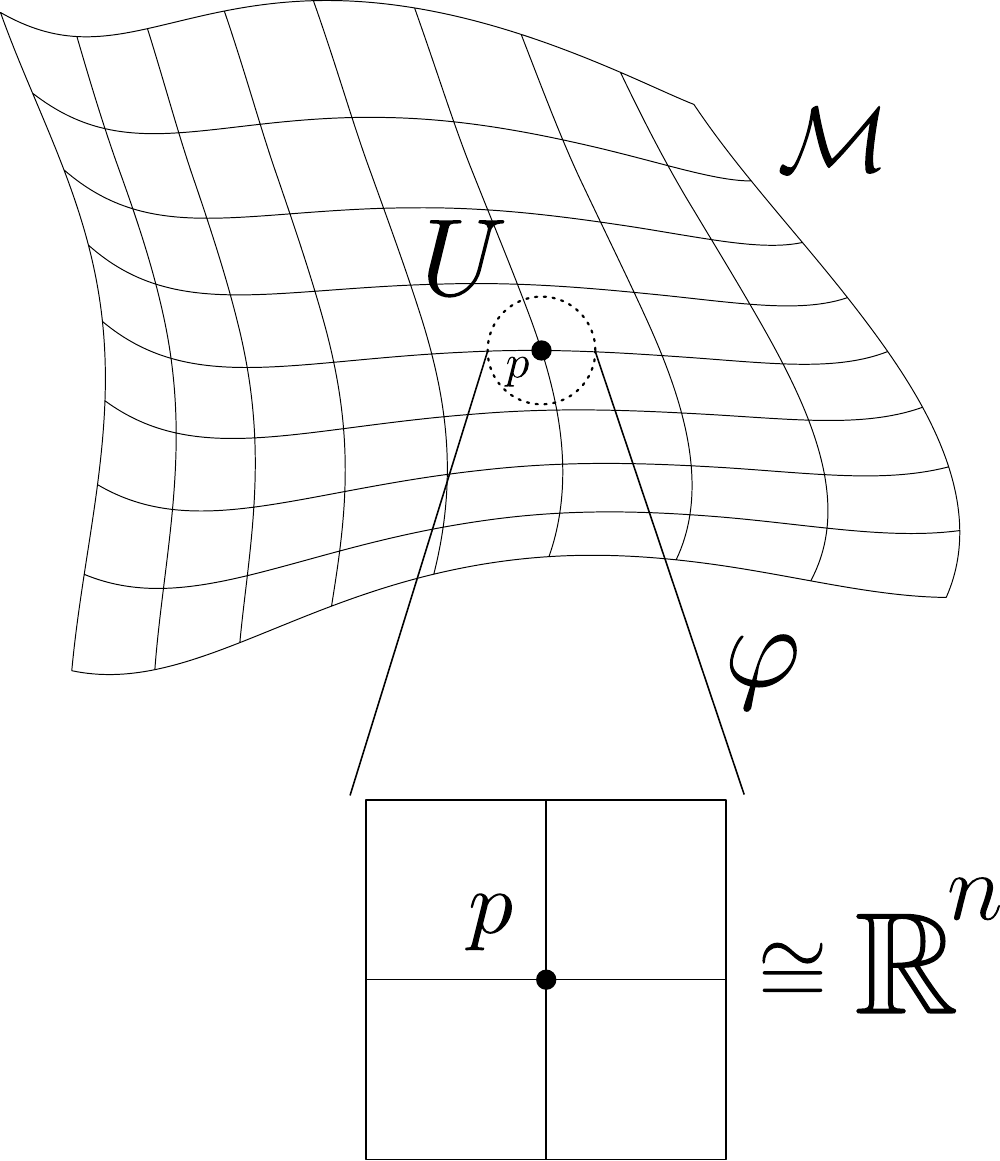}
  \caption{The chart $\phi$ is a (smooth) map from a neighborhood $U$ of a
    point $p$ on the general manifold $\mathcal{M}$ (not necessarily a surface) to an open
    subset of $\mathbb{R}^n$. }
  \label{fig:manifold}
\end{figure}
\subsection{Tangent space, bundle and frames}
\label{sec:tangent-space-frames}

We now consider at every point $p \in \mathcal{M}$, the tangent space
$T_p\mathcal{M}$ at that point, see Figure \ref{fig:tangent-space}. 
The collection of all such tangent spaces of $\mathcal{M}$ along
with their points of attachments viewed as a manifold itself is called a
tangent bundle \cite{leeIntroductionSmoothManifolds2012} \footnote{This is a
  fiber bundle with base space, where the fibers are given by the tangent spaces
at each point.}. We denote the total space
of the tangent bundle of $\mathcal{M}$ by $T\mathcal{M}$. An element of the tangent bundle 
$T\mathcal{M}$ is a pair $\left(p,u\right)$, where $p \in \mathcal{M}$ is the point of attachment 
of the tangent space on $\mathcal{M}$,
together with a tangent vector $u \in T_{p}\mathcal{M}$. We can express $p$ in coordinates
as e.g., $p = (A_w, A_n)$, and $u$ can be expressed via the components $(u_1,
u_2)$ of the vector $u$ expressed in some vector space basis of $T_p\mathcal{M}$. 
With the bundle structure follows the projection $\pi: T\mathcal{M} \rightarrow \mathcal{M}$. 
For each $(p,u)$, $\pi$ is just the projection onto the base point $p$, i.e., 
we ``forget'' about the vector $u$.

Since $\mathcal{M} \cong \mathbb{R}^{2}$, we have that for each 
$p \in \mathcal{M}$ that $T_p \mathcal{M} \cong T_p\mathbb{R}^2 \cong
\mathbb{R}^{2}$, and that  
$T\mathcal{M}  \cong T\mathbb{R}^{2} \cong \mathbb{R}^2 \times \mathbb{R}^{2}$. 
This means that $\mathrm{dim}\left( T\mathcal{M}  \right) = 4$. 

Consider now a general tangent vector field $V$ on $\mathcal{M}$, also called a
section of the bundle $T\mathcal{M}$. $V$ is a map $V: \mathcal{M} \mapsto T\mathcal{M}$, a choice of a vector $V_p \in
T_p\mathcal{M}$ at every point $p \in\mathcal{M}$. We are here assuming that this choice of
vector at each point is smooth in the sense that the vector components are
smooth functions on the manifold. Let $\mathbf{v}_i $ be a basis of
the tangent space $T_p\mathcal{M}$. We will use a bold font on general basis
vectors  to separate them from their coordinates. 
We can as usual expand any tangent vector $V_{p}$, $p\in \mathcal{M}$,
in the basis $\mathbf{v}_{i} $ as
\begin{equation}
  \label{eq:vector-basis-decomposition}
  V_{p} = v^{i}\left( p \right)\mathbf{v}_{i} \ \;,
\end{equation}
where $ v^{i} $ are the coordinates of $V_{p}$ with
respect to $\mathbf{v}_i$, which are functions of $p$. We use here and onwards the Einstein 
summation convention.  Similarly, we can expand a vector field $V$ using a set of sections
$\mathbf{s}_i$ as
\begin{equation}
  \label{eq:section-expansion}
  V \ = \ f^i\mathbf{s}_i \;,
\end{equation}
where $f^i$ are functions on $\mathcal{M}$.

We adopt the common convention that the basis
of tangent vectors at a point $p \in \mathcal{M}$ are directional derivatives acting on smooth
functions on the base space at that point
\cite{misnerGravitation1973,leeIntroductionSmoothManifolds2012}. A chart on some
open set $U \subset \mathcal{M}$ containing the point $p$, given e.g., by coordinate (functions)
$x^i(p) \equiv \left( A_w, A_n \right)$, gives a natural basis for the tangent space
$T_p\mathcal{M}$: the partial derivatives with respect to the coordinate
functions $x^{i}$ viewed as ``attached'' at $p$.

We introduce the notation
\begin{equation}
  \label{eq:local-basis-tangent-space}
  \left\{ \left. \frac{\partial}{\partial x^i}
    \right|_{p} \right\}_{i=1}^{n} \equiv \left\{ \left. \partial_{x^{i}}
    \right|_{p} \right\}_{i=1}^{n}\;,
\end{equation}
where $n$ is the dimension of the manifold. In the case
of $p = (A_w, A_n) \in \mathcal{M}$, we then have that
\begin{equation}
  \label{eq:local-basis-tangent-space-Aw-An}
  \left\{ \frac{\partial}{\partial A_w}, \frac{\partial}{\partial A_n} \right\} \equiv \left\{ \partial_w, \partial_n \right\}
\end{equation}
is a basis for the tangent space at each point $p$.
The partial derivatives act on smooth functions $f: \mathcal{M} \rightarrow \mathbb{R}$, which are simply functions that
take points on the manifold $\mathcal{M}$ as input. 
The total volumetric flow rate $Q = Q \left( A_w, A_n \right)$ is such a function.

We can now identify the thermodynamic velocities (\ref{eq:vw-therm}) and (\ref{eq:vn-therm}) as being
the basis $\left\{ \partial_w, \partial_n \right\}$ acting on the function $Q$; we have
``decoupled'' the vectors from the functions on which they act. The partial derivatives with respect to $A_w$
and $A_n$ at a point $p$, denoted by $\partial_w|_p$ and $\partial_n|_p$ respectively, acting
on the volumetric flow rate $Q$ define the
thermodynamic velocities. We have such a
derivation at each point $p \in \mathcal{M}$, so we can view $\left\{  \partial_w, \partial_n  \right\}$ as
coordinate vector fields on $\mathcal{M}$. These correspond to the sections
$\mathbf{s}_i$ in equation \eqref{eq:section-expansion}. In the same way, we
from now on identify any velocity with some tangent vector
acting on $Q$.  For instance, the pore velocity function $v$ can be identified with a tangent
vector field that has components $\left( S_{w}, S_{n} \right)$ in the basis $\left\{
\partial_w, \partial_n \right\}$, i.e.\ $S_w\partial_w+S_n\partial_n$. Upon acting on $Q$, we get the pore velocity function $v$. 

In the same way, we view the seepage velocities $v_i$ as
being defined by derivations acting on $Q$. In other words, we say that
there exists a basis
$\mathbf{e}_i$ of the tangent spaces of $\mathcal{M}$ that yield the seepage velocities upon acting on
$Q$,
\begin{equation}
  \label{eq:seepage-basis}
  \mathbf{e}_i \left( Q \right) \  \equiv \ v_i \;,
\end{equation}
where $i = w,n$.
The basis $\mathbf{e}_{i}$ is strictly speaking a frame, which 
means that the frame elements $\mathbf{e}_i$ could
be linearly dependent. 

In the following, we will
use the notation $v$, $\hat{v}_w$, $v_{n}$ etc. to signify the velocity functions, and use the 
notation $V$, $\partial_{i}$ and $\mathbf{e}_{i}$ for the vector fields associated with the velocity functions.

Note in particular that equation (\ref{eq:Q-euler-hom}) can be written as the action of a tangent
vector field on $Q$ which acts as the identity. We have a vector field $\Delta$ acting like
\begin{align}
  \label{eq:vf-Q}
  \Delta \left(  Q \right) \ =& \ A^i\partial_i \left( Q \right) = A_w \left(\frac{\partial Q}{ \partial
                           A_w}  \right)_{A_n} + A_n \left(\frac{\partial Q}{ \partial A_n}
                           \right)_{A_w} \;,
\end{align}
which by equation \eqref{eq:Q-euler-hom} is equal to $Q$ through the Euler theorem. 
Strictly speaking, we should be more careful with the notation: $A_w$ and $A_n$ as prefactors to
$\partial_w$ and $\partial_n$ are here coordinates on the ``fiber'', the tangent space. This
means that they are just the
components of a vector. Meanwhile, $A_w$, $A_n$ in $\partial_w$ and $\partial_n$ are the
coordinate {functions} on $\mathcal{M}$. We will not encounter problems by not
distinguishing them in this work, so we keep the notation as is for simplicity.

We have in this section shown how the velocities in Section \ref{sec:rea} can be
interpreted as objects on a tangent bundle with base space $\mathcal{M}$. When these
tangent vectors act on the function $Q$, we obtain the ordinary velocity
functions, which gives a number for each $p \in \mathcal{M}$. This simple fact is
the link to the ``classical'' geometric viewpoint we alluded to in Section \ref{sec:intro}. In what follows,
we will use this relation with ordinary numbers, and motivate the introduction
of affine spaces from the definition of the pore velocity $v$. We will then use the fact that we can relate
tangent vectors to points of an affine space (the tangent vector
spaces are actually affine spaces over themselves), and
show how this is helpful in the geometric interpretation presented in this work.

\subsection{Affine spaces of velocities, displacement- and tangent-vectors}
\label{sec:affine-space}

Before analyzing the co-moving velocity $v_m$ in terms of affine maps in Section \ref{sec:co-moving-affine-map}, 
we review some central points concerning affine spaces, the bundle structure, and the differences between the
differential-geometric and the affine descriptions presented in previous
sections.

Formally, an affine space
\cite{crampinApplicableDifferentialGeometry1987} is a set $\mathcal{A}$ of points together
with a vector space $\vv{\mathcal{A}}$, equipped with a map $\mathcal{A} \times \vv{\mathcal{A}} \rightarrow \mathcal{A}$. 
This map can be said to be the action of a vector $v \in \vv{\mathcal{A}}$ on a point $p \in \mathcal{A}$, acting as a
displacement to another point $p' \in \mathcal{A}$. The ``difference'' between two points
$p, p' \in \mathcal{A}$ can be identified with an element $v \in \vv{\mathcal{A}}$, which intuitively just
mean that the difference between two points can be identified with the vector
between them. We then have a space $\mathcal{A}$ of points, and a space $\vv{\mathcal{A}}$ of all
displacements between points of $\mathcal{A}$.

Coordinates on affine spaces entail a choice of an origin (a ``zero-vector'')
and linear basis with respect to this origin.
Consider an affine space $\mathcal{A}$ of dimension $n$, and let $o \in \mathcal{A}$ be a choice of
origin. Let $\left( \mathbf{e}_1, \mathbf{e}_2, \ldots, \mathbf{e}_n \right) \equiv \left\{ \mathbf{e}_i
\right\}$ be a choice of basis of $\vv{\mathcal{A}}$. Then, any point $p \in \mathcal{A}$ can be written as
\begin{equation}
  \label{eq:affine-coordinates}
  p=o + \left( p - o \right)=o + p^i\mathbf{e}_i\;,
\end{equation}
where $\left( p-o \right)$ is a vector since it is the difference of
two points, which we on the second line of
equation \eqref{eq:affine-coordinates} expanded in the basis $\mathbf{e}_i$ with components
$\left\{ p^{i} \right\}$. The components $\left\{ p^i \right\}$ are the
affine coordinates of the point $p$. A new choice of origin $o$ or 
basis $\left\{ \mathbf{e}_i \right\}$ specifies a new set of affine coordinates.
The choice of an origin and a linear basis with respect to this
origin is an affine frame or affine basis.

An affine map $f$ is a map between affine spaces that preserves the affine
structure. Any such map $f: \mathcal{A} \rightarrow
\mathcal{B}$ between affine spaces $\mathcal{A}$, $\mathcal{B}$, with associated vector spaces $\vv{\mathcal{A}}$ and
$\vv{\mathcal{B}}$ respectively, is defined by the property that for any two points $a,b
\in \mathcal{A}$, we have
\begin{equation}
  \label{eq:affine map-fundamental-property}
  f(a) - f(b) = \lambda \left( a-b \right) \;,
\end{equation}
where $\lambda$ is a linear map. Expressed equivalently, we have
\begin{equation}
  \label{eq:affine-map-def}
  f\left( p + u \right) \ = \ f \left( p \right) + \lambda u \;,
\end{equation}
where $p \in \mathcal{A}$ is a point, $u \in \vv{\mathcal{A}}$ is a vector, and $\lambda$ is a linear map. By
fixing points $o_1 \in \mathcal{A}$, and $o_2 \in \mathcal{B}$, a general affine map $f: \mathcal{A} \rightarrow \mathcal{B}$ can be
written in the form
\begin{equation}
  \label{eq:general-affine-map}
  f\left(p \right) \ = \ o_2 + \left( f (o_1) - o_{2} \right) + \lambda \left( p-o_{1} \right) \;,
\end{equation}
for $p\in \mathcal{A}$. Here, $ \left( f (o_1) - o_{2} \right) $ is a translation of $\mathcal{B}$
which only depends on $o_1$ and $o_2$, and $\lambda(p -o_1)$ is a linear map of the
vector $\left( p -o_1 \right) \in \vec{\mathcal{A}}$. At any point, we can form a vector
space and define some basis with respect to this point. We can for instance take
the derivative operators discussed in Section \ref{sec:tangent-space-frames} as a
basis for the vector space at this point. In this work, we can identify it with the tangent space
at that point. Thus, the
vector $\lambda\left(p-o_1 \right)$ can be treated as a tangent vector attached at
  $o_{1}$, obtained in coordinates by specifying affine coordinates for the
point $p$. A very important point is that by equation \eqref{eq:general-affine-map}, a
translation of the origin is also given by a vector, or equivalently a tangent
vector in this case.

A special case of an affine map is an invertible affine map from an affine
space $\mathcal{A}$ to itself, $f:\mathcal{A} \rightarrow \mathcal{A}$. Such a map is an affine
transformation of $\mathcal{A}$, and satisfies
\begin{equation}
  \label{eq:affine-transformation-general}
  f \left( p \right) \ = \ o + (f(o) - o) + \lambda \left( p - o \right)
\end{equation}
with $p \in \mathcal{A}$ and $o \in \mathcal{A}$ is taken as the origin, and $(f(o) - o)$ is a
translation.

For an affine combination of points $\left\{e_i
\right\}$ with coefficients $\alpha_i$, an affine map $f$ satisfies
\begin{align}
  \label{eq:affine-transformation}
  f \left(  \sum\displaylimits_{i=1}^n \alpha_i e_{i}\right) =&   \sum\displaylimits_{i=1}^n \alpha_i f\left( e_{i} \right) \ \;.
\end{align}

Let the tangent vectors introduced in the previous section act on $Q$,
such that we obtain the ordinary velocity functions. By the map given
in equations (\ref{eq:vw-transf}) and (\ref{eq:vn-transf}), we can rewrite the definition of the pore
velocity function $v$, equation \eqref{eqn10-5}, as
\begin{equation}
  \label{eq:vm-moving-origin}
  v= S_{w}\hat{v}_w + S_n\hat{v}_n
  =v_m + S_w \left(\hat{v}_w - v_m  \right) + S_n \left( \hat{v}_n - v_m \right)\;.
\end{equation}
Equation \eqref{eq:vm-moving-origin} expresses $v$ as an affine combination with $v_m$ singled out as a choice
of origin. Thus, we interpret the velocities $v$, $\hat{v}_i$, $v_i$ and $v_m$ as points of an
affine space $\mathcal{A}$, with an associated vector space $\vv{\mathcal{A}}$ of displacements. 
We view the space of velocities as affine since $v_m$ determines a
``moving origin'', $v_m = v_m \left( S_w \right)$.

Formally the velocities are points of $\mathcal{A}$, 
\begin{equation}
  \label{eq:points-vectors-in-spaces-affine}
  v, \hat{v}_i,  v_i , v_m \in \mathcal{A} \;, 
\end{equation}
whereas the velocity differences  $\left( \hat{v}_i - v_m \right)$  are {\it not\/} 
vectors in $\vv{\mathcal{A}}$. They
are just functions, giving a real number for each value of
the saturation $S_w$. 

Consider two pairs of points, for example the thermodynamic velocities $\left( \hat{v}_w, \hat{v}_{n}
\right)$ and the seepage velocities $\left( v_w, v_n \right)$ in $\mathcal{A}$. If we have a
map $g \in G$ for some group $G$ such that $g \left(\hat{v}_w, \hat{v}_n \right) =
\left( v_w, v_n \right)$, the group $G$ is said to act {2-transitively}
on $\mathcal{A}$. A prominent example of such a group is the group consisting of
translations and homotheties, or the group of dilations
\cite{bergerGeometry1994a}. These are examples of the affine transformations
just considered. The map $f$ in equation \eqref{eq:afffine-map-v} is exactly such a
2-transitive map. That $f$ acts $2$-transitively of $\left(\hat{v}_w, \hat{v}_n \right)$ means that if we know how one velocity is
mapped, the map of the other is known. This is exactly what is described by the relation defined in equations (\ref{eq:vw-transf}) and (\ref{eq:vn-transf}).

We now stress an important point regarding the relation between the ``classical''
and differential-geometric descriptions: the vector
space of displacements $\vv{\mathcal{A}}$ and the
tangent vector spaces $T_p\mathcal{M}$ at each point $p \in \mathcal{M}$ are formally not the same
spaces. However, they are isomorphic in the case
of $\mathcal{M} \cong \mathbb{R}^{2}$. The tangent spaces at each point of $\mathcal{M}$ can, 
in the case where we regard the
underlying space to be just $\mathcal{M} \cong \mathbb{R}^n$, be identified with each other by
translations. This is not possible in general; for a general
manifold, each tangent space must be viewed as distinct, as the concept of
simple displacements needs amending
\cite{crampinApplicableDifferentialGeometry1987}.
We note that in the
infinitesimal (tangent vector) case, the co-moving velocity $v_m$ is in general an
example of a particular type of section of a bundle, see \cref{sec:connections}. 

In Section \ref{sec:tangent-space-frames}, we defined the tangent vector spaces $T_p\mathcal{M}$, 
$p \in \mathcal{M}$, without endowing $\mathcal{M}$ itself 
with any particular structure. In fact, we could view $\mathcal{M}$ itself as an affine space. 
As an example of why this might be useful, consider the
case where we have some constant irreducible saturation in the two-phase flow
system. If the irreducible saturation is associated with some constant
non-vanishing flow rate, we have a constant term in our description of the areas
and the velocities that we have to take into account. The problem can then
be simplified if one could specify a new convenient origin in $\mathcal{M}$, for instance one corresponding to the irreducible saturation. 

Therefore, we seen that it can be useful to view $\mathcal{M}$ as not having a fixed
origin $O$. The latter was considered in \cite{pedersenParameterizationsImmiscibleTwophase2023a}. 
By specifying some origin $O$, one obtains a vector space structure. On the other hand, 
in order to refer to the relation between the choices of origins, one needs the 
affine structure. It turns out that in the case where we take the base space 
$\mathcal{M}$ to itself be an affine space, we can identify the tangent spaces at 
different points of $\mathcal{M}$ by translations of $\mathcal{M}$: given some
vector $u$, one can consider the translation or displacement $\tau_{u}: \mathcal{M} \rightarrow \mathcal{M}$ of all points of the affine space $\mathcal{M}$ by
this vector \cite{bergerGeometry1994a}. Note that this is a translation
  of all points of the space $\mathcal{M}$, and does not act as a derivation at a point as
  in the case of tangent vectors.  These translations are elements of the
vector space associated to $\mathcal{M}$ viewed as an affine space.

Let this associated vector space to $\mathcal{M}$ be denoted by $\vv{\mathcal{M}}$. We
note that $\vv{\mathcal{M}}$ can be identified with the ``vector space of areas'' from
earlier work \cite{pedersenParameterizationsImmiscibleTwophase2023a}. The
vector space $\vv{\mathcal{M}}$ associated to the affine space $\mathcal{M}$ can be 
viewed as containing the displacements between points of $\mathcal{M}$, just as with 
$\mathcal{A}$ and $\vv{\mathcal{A}}$ from earlier in this section. A tangent vector
$u \in T_p\mathcal{M}$ can be regarded as a tangent vector to a curve (which we
take to be just a line) $t \mapsto p + tu$ at the point $p \in \mathcal{M}$ \cite{crampinApplicableDifferentialGeometry1987}. Any displacement vector
$u^{\prime} \in \vv{\mathcal{M}}$ with the same direction as $u$ would give the same curve. 
If we consider the limit where the displacement given by $u^{\prime}$ goes to zero, we
see that we naturally have that we can let $u \in \vv{\mathcal{M}}$. Thus, we can view the 
tangent space at each point $p \in \mathcal{M}$ as a copy of $\vv{\mathcal{M}}$ 
attached to $p$. This identification between vectors of $\vv{\mathcal{M}}$ and vectors in $T_{p}\mathcal{M}$ at each $p \in \mathcal{M}$ is only possible due to the affine structure of $\mathcal{M}$, and it is important to note that this does not hold for general manifolds. 
This is so because there in general is no natural way of identifying vectors at different 
points of a manifold without introducing a \textit{connection} on the bundle \cite{leeIntroductionRiemannianManifolds2018, misnerGravitation1973}. Such a connection is extraneous to the
manifold itself. Note that the difference between the two is that the elements of
$\vv{\mathcal{M}}$ are, intuitively, ``detached'' from any point $p$.

To sum up, we only need a single space $\mathcal{M}$, whose displacements live
in the vector space $\vv{\mathcal{M}}$. We can either use the tangent vectors at each point to
describe the velocities at each point $p \in \mathcal{M}$, or we can let these tangent
vectors act on $Q$ and instead use the (signed) distances
between points of $\mathcal{M}$ as representing the displacements. 
This correspondence is possible due to the identification
$\mathcal{M} \cong \mathbb{R}^{2}$. In the latter case, we essentially do not use the manifold
structure of $\mathcal{M}$, and only treat it as the {linear} space $\mathbb{R}^{2}$.

\subsection{The saturation as a coordinate and parameter}
\label{sec:saturation-parameter-variable}

In the description of velocities as points in an affine space, equation
\eqref{eq:points-vectors-in-spaces-affine}, we have an important relation for
the space $\mathcal{M}$ of extensive variables: we can use the velocities to identify
``directions'' in $\mathcal{M}$. More explicitly, ratios of distances (the ``lengths'' of
the vectors in $\vv{\mathcal{M}}$) can be identified with points on an affine line $L \subset
\mathcal{M}$ through their functional values. We can specify points on this line either by
specifying a value of $S_w$,
or by specifying the values of the velocity differences. We will now clarify this point. 

The specific coordinates on $\mathcal{M}$ do not really matter
\cite{pedersenParameterizationsImmiscibleTwophase2023a}, so we specify points 
$p \in \mathcal{M}$ using the extensive areas, $p=\left( A_w, A_n \right) \in \mathcal{M}$. 
However, for practical reasons, it is often convenient to work with the coordinates
\cite{hansenRelationsSeepageVelocities2018} $(S_w,A_p)$, defined by
\begin{align}
  \label{eq:saturation-coordinates-Sw}
  A_p \ \equiv & \ A_w + A_n  \;, \\
  S_w \ \equiv & \ \frac{A_w}{A_w + A_n} = \frac{A_w}{A_{p}} \;. 
  \label{eq:saturation-coordinates-Ap} 
\end{align}
If we view  $A_p$ as fixed and constant, we only have a single
variable $S_w$. For each constant value $A_p = A_p^{*}$, $S_w$ parametrizes a line 
$L\subset \mathcal{M}$ running between $(A_w,A_n)=(0,A_p^*)$ and $(A_p^*,0)$. 
In these coordinates $\left( A_w, A_n \right)$, we have the
``trivial'' parametrization $\left( S_w A_p^{\ast}, \left(1 - S_w  \right) A_p^{\ast}\right)$. 

Since $\mathcal{M} \cong \mathbb{R}^{2}$, each $L$ (one for each value of $A_p^{\ast}$) 
can be seen as an affine subspace of $\mathcal{M}$ In terms of manifolds, $L$ is a 
sub-manifold of $\mathcal{M}$. 
The usage of the term ``affine subspace'' in this case is only due to our
identification of $\mathcal{M}$ with the real plane $\mathbb{R}^{2}$, 
viewed as a vector space itself. $S_w$ is in this context is called an 
affine coordinate on the line $L$. Moreover, $S_w$ is a parameter 
that specifies a point on the line $L$ defined by $A_w + A_n - A_p^{\ast} = 0$. 

The velocity functions are equivalent to one-dimensional maps
of the parameter $S_w$, which e.g., sends $S_w \mapsto \hat{v}_w \left( S_w \right) \in
L$. The relation between $S_{w}$ and the velocities are obtained by solving equation
\eqref{eq:v-both-definitions} for $S_w$, finding
\begin{equation}
  \label{eq:Sn-ratio}
  S_w \ = \ \frac{v-\hat{v}_n}{ \hat{v}_{w} - \hat{v}_{n}  } \ = \ \frac{v-v_n}{ v_{w} - v_{n} }
\end{equation}
where the velocity differences are simply the values of the corresponding functions. Thus, $S_w$ give the position of $v$ on the line
segment with $\hat{v}_w$ and $\hat{v}_{n}$ or $v_w$ and $v_{n}$ as endpoints for $v$.  

The view of $S_w$ as a parameter specifying a point on the line $L$
is quite useful for concrete computations. In fact, instead of letting $A_p$
equal a constant $A_p^*$, we can consider all relations ``modulo'' the scale
factor $A_p$, and work with the parameter
$S_w$ alone. By this, we mean that transformations
in the parameter $S_w$ are related to a (potentially continuous) family of lines $\left\{ L_{i}
\right\}$ in $\mathcal{M}$, where each line $L_i$ is given by a linear {inhomogeneous}
equation $aA_w + bA_n = c$, where $a$, $b$, $c$ are constants. This serves as
the entry point for continued work on the affine-geometric interpretation of the
system, and connects the affine relations in this work to projective
  geometry \cite{bergerGeometry1994a,
  richter-gebertPerspectivesProjectiveGeometry2011}. In this
context, where we can specify points on a line $L$ by using the ``dual''
intensive quantities to the extensive variables, the velocities $\left\{ v, \hat{v}_w, \hat{v}_n \right\}$,
or equivalently $\left\{ v, v_w, v_n \right\}$, can be called a type of
projective basis or projective frame \cite{gallierGeometricMethodsApplications2011, richter-gebertPerspectivesProjectiveGeometry2011}. A map of the
velocities sending $\hat{v}_i \mapsto v_i$ can in this context be said to be a map
defined on
the dual space of $\mathcal{M}$. What is meant by ``dual'' depends on the
context, but in this specific case, one is
referring to the projective dual of $\mathcal{M}$, denoted $\mathcal{M}^{\ast}$. 
This is simply the space where
each point $a \in \mathcal{M}^{\ast}$ represents a line in $\mathcal{M}$. The
velocities can then be seen as elements of $\mathcal{M}^{\ast}$, since they exactly specify
lines in $\mathcal{M}$. This can be seen by writing equation \eqref{eq:v-both-definitions} as
\begin{eqnarray}
  \label{eq:point-line-v}
  &&A_w \left( \hat{v}_w - v \right) + A_n \left( \hat{v}_n - v \right)\nonumber\\
  &=&   A_w \left( v_w - v \right) + A_n \left( v_n - v \right) = 0 \;.
\end{eqnarray}
In equation \eqref{eq:point-line-v}, $\left( A_w, A_n \right)$ specifies
points of $\mathcal{M}$, while the (ratio of the) velocities give the slope of
the line through the point $\left( A_w, A_n \right)$. In the special case that
$\mathcal{M} \cong \mathbb{R}^{2}$, this duality is trivial, however, this is the formal relation
between the extensive and intensive variables in the affine viewpoint. We will not need more specifics
about these spaces, and reserve this for future work.

\section{The co-moving velocity and affine maps}
\label{sec:co-moving-affine-map}

We will now investigate how the co-moving velocity $v_m$, first presented 
in equation (\ref{eq:co-moving-convex-comb}), can be described in terms of the two views
of the velocities presented in previous sections. As already mentioned in Section
\ref{sec:affine-space}, we have a natural identification between
the tangent spaces at each point of $\mathcal{M}$ and the vector space $\vv{\mathcal{M}}$ of
displacements of points of $\mathcal{M}$. From the discussion in the preceding sections,
we can work with either the distances between points given
by the differences $\left( v_i - v_m \right)$, or with tangent vectors at each point. We will start by using the
former description, where it is implicit that we have restricted ourselves to a
line $L \subset \mathcal{M}$ such that $S_w$ is a parameter along $L$, as discussed in the
previous section. We will then use the tangent vector description to write the
relations in terms of vector components, before simplifying the obtained
relations. The result will in the two cases be an expression for a function of
$v'=dv/dS_w$ and a vector field corresponding to the co-moving velocity $v_m$ 
respectively.

\subsection{$v_m$ from affine maps}
\label{sec:vm-from-affine-maps}

Let $f$ be an affine map. We now use use the property of affine maps in equation~\eqref{eq:affine-transformation}.
Comparing with equation \eqref{eq:v-both-definitions}, we see that the mapping
$\left\{ \hat{v}_i \right\} \mapsto \left\{ v_i \right\}$, which we call $f$, by
definition should satisfy
\begin{align}
  \label{eq:afffine-map-v}
  v \ = \ f \left( v \right) \ =& \ f\left( S_w \hat{v}_w + S_{n} \hat{v}_n
                                  \right) \nonumber  \\
  =& \ S_w f \left( \hat{v}_{w} \right) + S_n f \left( \hat{v}_{n} \right)
     \nonumber \\
  =& \ S_w v_w+ S_nv_n \;,
\end{align}
which holds since $S_w + S_n = 1$ at all times. Thus, $f$ can be seen as an
affine map $f: \hat{v}_i  \mapsto v_i$ leaving the convex combination $v$ invariant.

The details about the map $f$ depends on which interpretation we have for the
velocities. As expressed in the discussion around equation \eqref{eq:afffine-map-v}, $f$ as a map of the velocities is formally a map on $\mathcal{M}^{\ast}$, the space of lines in $\mathcal{M}$. However, since we can simply view the velocities as functions of $S_w$ only, $f$ is
simply a map of the one-dimensional number line $\mathbb{R}$. It is not important if this
number line is embedded in some higher dimensional space. We will call this
line $l$, the {image} of $S_{w} \in L$ under the velocity functions. This is what we will
take as the meaning of the map $f$ of the velocities: as a map of their
{functional values} on the line $l$. We will return to the case of $f$ acting on tangent
vectors, where the idea is exactly the same but expressed differently.

With the notion of an affine map $f$, we can revisit the right hand side of equation
\eqref{eq:vm-moving-origin}. The velocities $\left( \hat{v}_{n} + S_{n}v_m \right)$, $\left(
  \hat{v}_n + S_w v_m \right)$ are in this case not velocity differences; they
are expressions of a particular affine map called
a {homothety}, see Section  \ref{sec:homotheties}. In
fact, $v$ itself can be written as a homothety. To see this, we rewrite 
equation~\eqref{eq:v-both-definitions} as
\begin{equation}
  \label{eq:v-as-a-homothety}
  v=\hat{v}_w + S_n \left( \hat{v}_n - \hat{v}_w \right)=v_w + S_n \left( v_n - v_w \right) \;,
\end{equation}
where the middle and third expressions respectively are homotheties of ratio $S_n$
with centers $v_w$ and $\hat{v}_w$
\cite{bergerGeometry1994a}. 

Consider $\left\{ \hat{v}_i \right\}$ and $\left\{ v_i\right\}$ as points in two
affine spaces $\mathcal{A}$, $\mathcal{B}$ with associated vector spaces $\vec{\mathcal{A}}$ and
$\vec{\mathcal{B}}$ respectively. Using equation~\eqref{eq:vm-derivative-form} and equation \eqref{eq:afffine-map-v}, we
have that
\begin{align}
  \label{eq:affine-map-derivative-form}
  f \left( \hat{v}_w \right) - f \left( \hat{v}_n \right)  \ =& \  v_w - v_n \nonumber
  \\
  =& \ \frac{d v}{d S_w} - v_m \;.
\end{align}
The velocity difference
$(v_w - v_n)$, where $v_i = f(\hat{v}_i)$, is then equivalent to a
 linear map $\lambda$ of $\left( \hat{v}_w - \hat{v}_{n} \right)$
according to Section  \ref{sec:affine-space}.  In writing, $\left( f (\hat{v}_w) - f (\hat{v}_n)
\right) = (v_w - v_n)$, we can specify a choice of origin in $\mathcal{A}$ and $\mathcal{B}$. We choose the origins
$o \in \mathcal{A}$ and $p \in \mathcal{B}$, and use equation \eqref{eq:general-affine-map} and
equation \eqref{eq:affine-map-derivative-form} to write 

\begin{align}
  \label{eq:velocity-relation-affine-map}
  f (\hat{v}_w) = f (\hat{v}_n) + \lambda (\hat{v}_w - \hat{v}_n)  \;,
\end{align}
which from Section \ref{sec:affine-space} is equivalent to
\begin{align}
  \label{eq:equivalent-affine-map}
f (o + u) \ =& \ p + (f(o) - p) + \lambda u \nonumber \\
  =& \ p  + \lambda u \;,
\end{align}
for some vector $u$, and where we let $f(o) \equiv p$. We can then set
$f \left( o + u \right) = f \left( \hat{v}_w \right)$, the new origin $p = f \left( \hat{v}_{n}
\right)$, $\lambda u =  \left( v^{\prime} - v_m \right)$. The point is that the affine map
$f$ also moves the origins of the velocities.

As before, we can associate the vector $u$
with its (Euclidean) length of the distance between points on the line $l$. Thus, the
meaning of equation \eqref{eq:equivalent-affine-map} is simply that a velocity defined by
the distance $u$ from some origin $o$ is mapped to a new origin $p$ and a
linear map of the distance $u$. Even if we a priori have no preferred way of
defining such an origin or vector $u$, the map $f$ suggests that the origin
should move, and the distance $u$ from the origin is scaled by $\lambda$.

We are now ready for a simple yet important result. Comparing equation \eqref{eq:affine-map-derivative-form} to the definition in equation~\eqref{eq:affine map-fundamental-property}, we see that we can write
\begin{align}
  \label{eq:affine-map-of-v-prime}
  f \left( \hat{v}_w \right) - f \left( \hat{v}_n \right) \ =& \ \lambda \left( \hat{v}_w - \hat{v}_n \right) \nonumber \\
  =& \ \lambda \left( v^{\prime} \right) \ = \ v^{\prime} - v_m \;,
\end{align}
where $\lambda$ is a linear map. In one dimension, the only linear maps are
multiplication by a scalar, so $\lambda$ is just a number. Thus, we have
\begin{equation}
  \label{eq:vm-v-prime-affine}
  v_m \ = \ \left( 1 - \lambda \right) v^{\prime} \;.
\end{equation}
Comparing equation \eqref{eq:vm-v-prime-affine} to equation \eqref{eq:vm-constitutive}, we
see that we can identify $(1 - \lambda) \equiv b$. However, the term $a v_0$ in equation \eqref{eq:vm-constitutive} does not
appear from considering an affine map $f$ in this way. We will see that we get
the same result when treating the
velocities as tangent vector fields in \cref{sec:vm-tangent-vector-field}.

\subsection{Homotheties and irreducible capillary flow }
\label{sec:homotheties}

Intuitively, equation \eqref{eq:v-as-a-homothety} means that $v$ is the point located
at a fraction $S_n$ along the line segment between $\hat{v}_w$ and $\hat{v}_n$ in
$\mathcal{A}$. Thus, if either $\hat{v}_w$ or $\hat{v}_n$ (or both) were to change while
$v$ was kept fixed, $S_n$ would also change, and hence also $S_w=1-S_n$. 
Thus, any change in one of the thermodynamic velocities
is accompanied by an equal and opposite change in the other thermodynamic velocity. This
is the relation between the middle and last expression in equation \eqref{eq:v-as-a-homothety},
and also in equation \eqref{eq:afffine-map-v}. In fact, affine maps are the only maps
that ``commute'' with the saturation $S_n$ in this way, which we see is
the defining property which we use to introduce $v_m$ through
\cref{eq:vw-transf,eq:vn-transf}.

Let $\tau_{u}$ be a translation of $\mathcal{A}$ by the vector $u$ and let $h_{s,
  \lambda}$ be a homothety of center $s$ and ratio $\lambda$. A composition of $\tau_{u}$
and $h_{s, \lambda}$, since the set of translations and homotheties of an affine space
forms a group \cite{bergerGeometry1994a}, is again a homothety
$\tilde{h}_{\tilde{s}, \lambda}$ of center $\tilde{s}$. Explicitly, if $h_{s, \lambda}: p \mapsto
s + \lambda \left( p-s \right)$, and $\tau_{u}: p \mapsto p + u$ with $u$ a
vector, we have
\begin{equation}
  \label{eq:composition-translation-homothety}
  \tau_{u} \circ h_{s, \lambda}: p \mapsto s + u + \lambda \left( p-s \right) \;.
\end{equation}
We can rewrite equation \eqref{eq:composition-translation-homothety} as a new
homothety $\tilde{h}$ with respect to a point $\tilde{s}$ and the same ratio $\lambda$
as
\begin{equation}
  \label{eq:composition-translation-homothety-better} 
  \tau_{u} \circ h_{s, \lambda}: p \mapsto  s + \frac{u}{1-\lambda} + \lambda \left( p - \left( s +
      \frac{u}{1-\lambda} \right) \right) \;.
\end{equation}
If we define
\begin{equation}
  \label{eq:homothetic-center}
  \tilde{s} = s + \frac{u}{1- \lambda} \;,
\end{equation}
we can define $\tau_{u} \circ h_{s, \lambda}\equiv \tilde{h}_{\tilde{s}, \lambda}$ and write equation
\eqref{eq:composition-translation-homothety-better} as
\begin{equation}
  \label{eq:homothety-new-center}
  \tilde{h}_{\tilde{s}, \lambda} : p \mapsto \tilde{s} + \lambda \left( p - \tilde{s} \right) \;.
\end{equation}

The formalism in terms of homotheties as defined above can be applied directly to the system studied in Section 7.3 of
Reference \cite{hansenRelationsSeepageVelocities2018}.  This system consist of $N$ capillary fibers in parallel, of
which $N_s$ have smaller cross section $a_s$ and the rest, $N_l=N-N_s$ have a larger cross section $a_l$. We assume the
smaller cross section is so small than only the wetting fluid can enter these capillaries.  Each capillary is filled with 
either wetting or non-wetting only. 
The wetting pore area is then $A_w = A_s + A_{lw}$ where $A_s=N_s a_s$ and $A_{lw}$ is the area of the large capillaries that
are filled with wetting fluid.  This means that the system has an irreducible saturation given by $S_{w,i}=A_s/A_p$. Hence,
the wetting area is given by $A_w=A_pS_{w,i} + A_p\left( S_w - S_{w,i} \right)$. The non-wetting saturation is given by
$S_n=1-S_w$.  We denote the velocity of
the non-wetting fluid by $v_n$, and the velocity of the wetting fluid in the
small capillaries $v_{sw}$ and in the large capillaries by $v_{lw}$. 
The average flow velocities through the capillary fiber bundle is then
\begin{equation}
  \label{eq:v-irreducible-flow-example}
  v \ = \ S_{w,i} v_{sw} + \left( S_w - S_{w,i} \right)v_{lw} + S_n v_n\;.
\end{equation}
We may now interpret the velocities as points in a space $\mathcal{A}$. 
We combine equations (\ref{eq-ah-1}) and  (\ref{eq:v-irreducible-flow-example}) to find
\begin{equation}
  \label{eq:irreducible-wetting-example-swvw}
  S_w v_w \ = \ S_{w,i} v_{sw} + \left( S_w - S_{w,i} \right)v_{lw} \;.
\end{equation}
We express $v_w$ by dividing the left hand side of equation
\eqref{eq:irreducible-wetting-example-swvw} by $S_w$, and insert this into
equation~\eqref{eq:v-as-a-homothety} to obtain
\begin{align}
  \label{eq:v-substituted-irreducible-example}
  v \ =& \ v_w + S_n \left( v_n - v_w \right) \nonumber \\
  =& \ \left[ v_{lw} + \frac{S_{w,i}}{S_w}\left( v_{sw} - v_{lw} \right) \right]
     \nonumber \\
     +&
   S_n \left[ v_n - \left( v_{lw} + \frac{S_{w,i}}{S_w}\left( v_{sw} - v_{lw}
  \right) \right) \right] \nonumber \\
  =& \ v_{lw} + \frac{S_{w,i}}{1-S_n}\left( v_{sw} - v_{lw} \right) 
     \nonumber \\
     +&
   S_n \left[ v_n - \left( v_{lw} + \frac{S_{w,i}}{1-S_n}\left( v_{sw} - v_{lw}
   \right) \right) \right] \;.
\end{align}
Comparing equation
\eqref{eq:v-substituted-irreducible-example} to 
equations (\ref{eq:composition-translation-homothety-better}) and (\ref{eq:homothetic-center}), we see
that we have defined a composition of a homothety of ratio $S_n$ of the point $v_n$ with
respect to the center $v_{lw}$ and the translation of the point $v_{lw}$ by the
constant vector $S_{w,i} \left( v_{sw} -v_{lw} \right)$.  We can thus 
identify it with a translation of the origin from
$S_{w,i}v_{lw}$ to $S_{w,i}v_{sw}$. We may then rewrite equation
\eqref{eq:v-substituted-irreducible-example} one last time as
\begin{equation}
  \label{eq:v-irreducible-flow-rewritten}
  v  \ = \ v_{lw} - v_m + S_n \left(  v_n - (v_{lw} - v_m) \right) \;,
\end{equation}
where we have identified 
\begin{equation}
\label{eq:ah-500}
v_m = \frac{S_{w,i}}{S_w} \left( v_{lw} - v_{sw} \right)\;.
\end{equation}
Comparing equation \eqref{eq:v-irreducible-flow-rewritten} to equation
\eqref{eq:homothety-new-center}, we get that $\tilde{s} = v_{lw} - v_m$, meaning it
can be viewed as a translation of the homothetic center $v_{lw}$. $v_m$ is
exactly the translation vector of the homothetic center in the space of velocities.

We find from equation (\ref{eq:v-irreducible-flow-example}) that
\begin{equation}
\label{eq:ah-501}
v'=\frac{dv}{dS_w}=v_{lw}-v_n\;.
\end{equation}
Hence, $v_m$ in this system is {\it not\/} on the form suggested by equation 
(\ref{eq:vm-v-prime-affine}). The reason for this is that there is no mechanism in the
 capillary fiber bundle system to generate an equilibrium thermodynamics as the fibers are non-interacting. 

We have in \cref{sec:vm-from-affine-maps} related $v_m$ to the affine map $f$,
with the result that \cref{eq:vm-v-prime-affine}
is linear in $v^{\prime}$. \Cref{eq:affine-map-of-v-prime} means that $f$ acts the
{same way} on both thermodynamic velocities. This restriction is what
gives us~\cref{eq:vm-v-prime-affine}. When the constituent
subsystems do not interact with each other as in the capillary fiber bundle example, which in general can
happen in sub-regions of the saturation range, we see that we do not
have a single map $f$ acting as in \cref{eq:affine-map-of-v-prime}. However, the
velocity $v$ itself can be expressed in terms of a homothety, which is affine.
This means that~\cref{eq:vm-v-prime-affine} is not correct in this case. Solving
\cref{eq:ah-500} for $v_{lw}$ and inserting into \cref{eq:ah-500} gives us
\begin{equation}
  \label{eq:vm-non-trivial-fiber-bundle}
  v_m = \frac{S_{w,i}}{S_w} v^{\prime} + \frac{S_{w,i}}{S_w} \left( v_n - v_{sw} \right) \;,
\end{equation}
which contains a linear transformation in $v^{\prime}$ \footnote{The map is
  non-linear in $S_w$, but appears as a multiplicative factor of $v^{\prime}$, hence
  the map of $v^{\prime}$ is linear.} and a translation term which moves the origin.
In \cref{sec:connections}, we will see some solutions for describing this geometrically.

\subsection{$v_m$ as a tangent vector field}
\label{sec:vm-tangent-vector-field}

We now turn to the interpretation of velocities as tangent vectors in the
tangent spaces of $\mathcal{M}$, where $\mathcal{M}$ is again viewed as a manifold. 
We will exploit the fact that $\mathcal{M}  \cong \mathbb{R}^{2}$ to circumvent a discussion of
connections (see Section \ref{sec:connections}   ) and mathematical fiber
bundles (see). We will simply say that we are
able to choose an origin $o$ in each tangent space that may depend on the
point $p \in \mathcal{M}$. We regard vectors in the tangent space $T_p \mathcal{M}$ to be
attached at the point $o \in T_p\mathcal{M}$. This origin is given by {some} section $s$ of $T\mathcal{M}$ which we assume to
be {non-vanishing} for the domain in $\mathcal{M}$ we are considering. This means
that the vector field itself has no singular points. A section of a bundle exists independently of a representation in terms of
coordinates, so there is no intrinsic way of defining coordinates for
a section unless more structure is provided.

We can encode an ``indeterminate'' origin in the tangent spaces of the bundle
$T\mathcal{M}$ by letting the origin of the tangent spaces be given by a section $s_0$.
This gives us an affine bundle \cite{saundersGeometryJetBundles1989}, where the
fibers are now related by affine maps.
We cannot use the choice $s_0$ to define the coordinates of the section $s_{0}$
itself; the choice of $s_{0}$ is rather a part of
 the choice of affine frame generalized to the bundle (see
 \cref{sec:affine-space}), which allows us to define coordinates for vectors \footnote{An application of this
  formalism is seen in mechanics, see \cite{desaxceAffineTensorsMechanics2012}.
  We follow the same reasoning here.}.

For now, we disregard $s_0$, and consider a single tangent space
$T_p\mathcal{M}$. Recall that $T_p\mathcal{M}$ is itself an affine space, denoted $\mathcal{A}_{p}$ with an
associated vector space $\vv{\mathcal{A}}_{p}$. We consider the origin of $\vv{\mathcal{A}}_p$ as a point $\hat{o} \in \mathcal{A}_p$.
Let another choice of origin be $o$. A choice of $o$ in each fiber is determined
by a section $s$. In each tangent space, we can then identify a {vector} $\vv{\left( o\hat{o} \right)}_{p}
  = u \in \vv{\mathcal{M}}_{p}$
between the points $\hat{o}, o \in \mathcal{A}_{p}$, and {this} vector can be decomposed into components. We
have a choice of such a vector in each tangent space, given by the section
$s$. Thus, we can associate the section $s$ to a vector field that we can
describe using vector components. We rename this field from $s$ to $V_m$, to
make the analogy clear. Note that this is exactly what is implied by the right hand side of equation \eqref{eq:vm-moving-origin}. 

We use the index $\alpha = w,n,$ to label which velocity we are referring to. For
each $\alpha$, a vector is given by two components, because $\mathrm{dim} \left(
  T_p \mathcal{M} \right) = 2$. Since we here view each $T_p \mathcal{M}$ as an affine space, every
vector is defined with respect to some choice of origin which in general is a
function of the point $p \in \mathcal{M}$. As done before, we use the notation $\mathcal{A}$ for the
affine space of points corresponding to $T_p\mathcal{M}$, and $\vv{\mathcal{M}}$ for the associated
vector space.

We label the velocities viewed as points of the affine space $\mathcal{A}_p$ by a left
superscript ${}^{p}\left( \cdot \right)$. Thus, the thermodynamic velocities, denoted  ${}^{p}\hat{v}_{\alpha}$, and the seepage
velocities,  ${}^pv_{\alpha}$, which we stress are not functions but abstract
{points} of $T_p\mathcal{M}  = \mathcal{A}$, are then expressed as
\begin{align}
  \label{eq:therm-affine}
{}^{p}\hat{v}_{\alpha} \ = \ \hat{o}_{\alpha} + \vec{\hat{v}}_{\alpha} \;, \\
  {}^{p}v_{\alpha} \ = \ o_{\alpha} + \vec{v}_{\alpha} \;, \label{eq:seepage-affine}
\end{align}
where $\hat{o}_{\alpha}, o_{\alpha} \in \mathcal{A}$, $\vec{\hat{v}}_{\alpha}, \vec{v}_{\alpha} \in \vv{\mathcal{A}}$. We here
regard the points and velocities corresponding to the thermodynamic- and seepage
velocities to belong to the same affine- and vector space, which simplifies the
notation in Section \ref{sec:co-moving-affine-map}.

In the notation introduced above, we can write the relations in  equations (\ref{eq:vw-transf}) and (\ref{eq:vn-transf} as
\begin{align}
  \label{eq:vi-affine-rewrite}
  {}^{p}\hat{v}_{\alpha} - {}^{p}v_{\alpha} \ = & \  (\hat{o}_{\alpha} +  \vec{\hat{v}}_{\alpha}) - \left(o_{\alpha} +
    \vec{v}_\alpha  \right) \nonumber  \\ 
  =& \ \left(  \hat{o}_{\alpha} - o_{\alpha}\right) + \left( \vec{\hat{v}}_{\alpha} - \vec{v}_{\alpha}
     \right) \nonumber \\
  =& \ O_{\alpha}^j \mathbf{e}_{j, \alpha} + \lambda_{\alpha}^j \mathbf{e}_{j, \alpha} \nonumber \\
\equiv& \ v_{\alpha}^j \mathbf{e}_{j, \alpha} \;,
\end{align}
where the index $j$ runs over the dimension of $\mathcal{A}$, $\mathrm{dim} \left( \mathcal{A}
\right)= 2$, and $v_{\alpha}^j = \left( O_{\alpha}^j + \lambda_{\alpha}^j \right)$. $\lambda_{\alpha}^j$
are the components of the tangent vectors in $\vv{\mathcal{A}}$. Thus, we see that by
introducing a shift in the origin, the new components are {linear
  inhomogeneous} functions (in other words {affine} functions) of the
components $\lambda_{\alpha}^j$.\footnote{Affine functions of the components of the vectors
  is central in the definition of affine bundles, see e.g. \cite{sardanashvilyAdvancedDifferentialGeometry2013}.}

In equation \eqref{eq:vi-affine-rewrite}, we expanded $\left(  \hat{o}_{\alpha} -
  o_{\alpha}\right)$ and $\left( \vec{\hat{v}}_{\alpha} - \vec{v}_{\alpha}
\right)$ in the same basis $\mathbf{e}_{\alpha}^{j}$. In particular, the basis is not
dependent on which velocity we are referring to, as the basis is the same for
all $\alpha$. Therefore, we have $\mathbf{e}_{j, \alpha} = \mathbf{e}_j$, i.e. we drop the
index $\alpha$.

The last line in equation \eqref{eq:vi-affine-rewrite} then defines {two} co-moving velocities,
\begin{align}
  \label{eq:two-vm-w}
  {}^{p}\hat{v}_w -  {}^{p}v_w \ =& \ v_w^j \mathbf{e}_{j} \ \equiv \ S_n\vv{v_{m}^w} \;, \\
  {}^{p}\hat{v}_n - {}^{p}v_n \ =& \ v_n^j \mathbf{e}_{j}  \ \equiv \ -S_w\vv{v_{m}^n} \label{eq:two-vm-n} \;.
\end{align}
The quantities $\vv{v_m^{\alpha}}$ are written as vectors because they are defined as
the difference between two points, hence vectors. In Section \ref{sec:rea}, invariance of $v$ requires that $\vv{v_m^w} = \vv{v_m^n} \equiv \vv{v_m}$.
This places restrictions upon the coefficients $v_{\alpha}^j$.

We now adopt the view in equation \eqref{eq:afffine-map-v}, namely that the mapping
in equation \eqref{eq:secondary-transf} is given by an affine map $f$, or in this
case: an affine transformation. This view
presents no new difficulties, and just means that we view the components in equation
\eqref{eq:vi-affine-rewrite} as being related by the map $f$. From Section 
\ref{sec:affine-space}, we then have that
\begin{align}
  \label{eq:affine-map-explicit-linear}
  \vec{v}_{\alpha} \ =& \ \bar{f} \left( \vec{\hat{v}}_{\alpha} \right) \;, \\
  o_{\alpha} \ =& \ f(\hat{o}_{\alpha}) \label{eq:affine-map-explicit-origin} \;,
\end{align}
where $\bar{f}$ is the linear part of the affine map $f$. If the components
of the linear part $\vec{\hat{v}}_{\alpha}$ of ${}^{p}\hat{v}_{\alpha}$ (same for the seepage
velocities) with respect to the basis $\mathbf{e}_j$ are $x_{\alpha}^j$, then the
components of $\vec{v}_{\alpha}$ are related to $x_{\alpha}^j$ by a linear transformation,
$x_{\alpha}^j \mapsto x_{\alpha}^i\bar{\lambda}_{i,\alpha}^{j}$. Since we in equation \eqref{eq:afffine-map-v}
only use a single map $f$, the linear transformation is equal for $\alpha = w,n$, so
we can drop the index $\alpha$ on the matrix representation of the linear
transformation. The assumption that the mapping in equation
  \eqref{eq:secondary-transf} is given by a single map $f$ is the simplest
  choice we can make. If we allowed for a pair of maps, meaning that $f \mapsto
  f_{\alpha}$, the transformation would not be affine. This would also imply that we have
  {two} thermodynamic velocities, meaning $\vv{v_m^w} \neq \vv{v_m^n}$, which makes us
  unable to define a single $\vv{v_m}$ and hence a single $v_m$.\footnote{In
    \cref{eq:vi-affine-rewrite}, this assumption also means that the origins for
    $v_w$ and $v_n$ are taken to be the same.}
    
From the above, we can now rewrite equation
\eqref{eq:vi-affine-rewrite} as 
\begin{align}
  \label{eq:vi-affine-rewrite-affine-map}
  \hat{v}_\alpha - v_\alpha \ = & \ \left(  \hat{o}_\alpha - o_\alpha\right) + \left( \vec{\hat{v}}_\alpha - \vec{v}_\alpha
                        \right) \nonumber \\ 
  =& \ \left(  \hat{o}_\alpha - f( \hat{o})_\alpha\right) + \left( \vec{\hat{v}}_{\alpha} -
     \bar{f}\left( \vec{v}_{\alpha} \right) \right) \nonumber \\
  =&  \ O_{\alpha}^j \mathbf{e}_{j} + \left( x^j_\alpha -
     x^i_\alpha\bar{\lambda}_{i}^{j} \right)\mathbf{e}_j \nonumber \\
  =&  \ \left(O^j_{\alpha} + x^i_\alpha \left( I_i^{j} - \bar{\lambda}_{i}^{j}\right)
      \right)\mathbf{e}_j \;.
\end{align}
where $I_i^{j}$ is the identity matrix, 
$x_{\alpha}^j$ are the components of $\vec{\hat{v}}_{\alpha}$ in the basis $\mathbf{e}_j$, and
$\bar{\lambda}_{i}^{j}$ is the matrix-representation of the linear transformation
$\bar{f}$.
Using equation \eqref{eq:vi-affine-rewrite-affine-map} and equations (\ref{eq:vi-affine-rewrite}) --- (\ref{eq:two-vm-n}), we can then write the
difference $\left( {}^{p}v_w - {}^{p}v_n \right)$ as equations (\ref{eq:two-vm-w} and (\ref{eq:two-vm-n})\footnote{Note that if we had allowed for unequal origins, the matrices in equation
  \eqref{eq:vw-minus-vn-real-stuff} would have the index $\alpha$ (i.e. they would be
  tensors), and there would be an additional term due to potentially different
  choices of origin for $v_w$ and $v_n$.}
\begin{align}
  \label{eq:vw-minus-vn-real-stuff}
  {}^{p}v_w - {}^{p}v_n \ =& \ \left( v_w^j - v_n^j \right)\mathbf{e}_{j}  \nonumber \\
  =& \ \Big[ \left( x^j_w - x^j_n
       \right) 
   + \left(x^i_{n} \bar{\lambda}^j_{i}-  x^i_w \bar{\lambda}^j_{i}  \right)\Big] \mathbf{e}_{j} \nonumber \\
 =& \ \Big[ \left( x^j_w - x^j_n
       \right) 
  + \left(x^i_{n} -  x^i_w \right) \bar{\lambda}^j_{i}  \Big] \mathbf{e}_{j} \nonumber \\
  =& \left( x^{i}_w - x^{i}_n\right)\left(  I_i^{j} - \bar{\lambda}_{i}^{j} \right)\mathbf{e}_{j}
\end{align}
To relate equation \eqref{eq:vw-minus-vn-real-stuff} to $v_m$, we need to make some restrictions. We choose coordinates
$\left( S_w, A_p \right)$ on $\mathcal{M}$, which induces the coordinate basis $\left(
  \partial_{S_w},
  \partial_{A_{p}} \right)$ on the tangent space. In general, both vectors enter into
equation \eqref{eq:vw-minus-vn-real-stuff}. We can now either set $A_p = A_p^{\ast}$
where $A_p^{\ast}$ is a
constant (which essentially means that we restrict to a subspace or ``sub-manifold'' of $\mathcal{M}$), or
consider the extensive variables on $\mathcal{M}$ up to a common factor of
$\lambda$, where $\lambda$ need {not} be constant (we could here e.g. set $\lambda = A_p$). The two possibilities give us two
potential definitions of the saturation, which we could label $S_w^{A_p^{\ast}}$
and $S_w^{\lambda}$. The former definition is the most straightforward, where all
quantities are seen in relation
to a ``absolute'' total area. The
latter possibility is related to {projective spaces}, which is outside
the scope of this work. However, no matter which of the two possibilities is invoked, the
basis $\mathbf{e}_j$ reduces to a single element, and we label this single
element simply by $\partial_{S_w}$ as before. The fourth line in
equation~\eqref{eq:vw-minus-vn-real-stuff} is then trivial, and can be simplified to
\begin{align}
  \label{eq:vw-minu-vn-simpler}
  {}^{p}v_w-  {}^{p}v_n \ =& \ \tilde{\gamma} \partial_{S_w}  \nonumber \\
  =& \ \left( 1 - \gamma \right) \partial_{S_w} 
\end{align}
equation \eqref{eq:vw-minu-vn-simpler} can, upon acting on the function $Q$, be identified with $\left( v^{\prime} - v_m
\right)$, so that $\gamma \partial_{S_w} \equiv \vv{v}_m$ where $\gamma$ is a function on $\mathcal{M}$. Equation
\eqref{eq:vw-minu-vn-simpler} has the same content as equation
\eqref{eq:vm-v-prime-affine}, only described in terms of tangent vectors.
Moreover, as in equation~\eqref{eq:vm-v-prime-affine}, we cannot explicitly get a term
corresponding to the constant $a v_0$ in equation \eqref{eq:vm-constitutive}. We
will return to this and its generalization in \cref{sec:connections}.

\section{Discussion and connection to further formalisms}
\label{sec:discussion-related}

The two ways of viewing the velocities discussed in this work might seem
almost equivalent, but the two approaches represent very different views of the
base space and the velocities. In viewing the velocities simply as ``points on a
line'' in Section \ref{sec:affine-space}, the velocities are then examples of
{homogeneous coordinates} \cite{bergerGeometry1994a,
  spivakComprehensiveIntroductionDifferential1999} on projective spaces. These coordinates can be
interpreted as labels for points along the real number line (in our case), which
we are free to regard as the function values corresponding to the velocities.
The co-moving velocity is in this context simply another point on the number
line. This way of viewing the velocities allow for working with concrete
numbers. In viewing the velocities as points of an affine space attached to each
point of the base space of extensive variables, we loose the ``tangibility'' of
the ``classical'' method. However, the language of bundles and manifolds underpins
investigations into the geometry of thermodynamics.

As a sidenote, we see from the considerations in Section
\ref{sec:co-moving-affine-map} that
we cannot get an isolated constant term $a v_0$ as in the
phenomenological constitutive relation in equation \eqref{eq:vm-constitutive} in the
framework presented here.\footnote{Strictly speaking, it {can} be done if the
  factor in front of $v^{\prime}$ and $\partial_{S_w}$ in equation \eqref{eq:vm-v-prime-affine}
  and equation \eqref{eq:vw-minu-vn-simpler} contains a term that cancels $v^{\prime}$
  exactly as $v^{\prime} \rightarrow 0$, i.e. a term $\sim (v^{\prime})^{-1}$. However, such a term would
  not correspond to any well-defined vector field, or would require knowing the
function $Q$ itself.} This conclusion follows from the
observation that the map in equation \eqref{eq:secondary-transf} is given by a single
affine transformation, the affine map $f$. 
The term $av_0$ was first obtained \cite{hansenRelationsSeepageVelocities2018} 
from fits of experimental data. A physical interpretation of it was then presented in
\cite{hansen2024linearity}: we have from equations (\ref{eq:vm-derivative-form}) and (\ref{eq:vm-constitutive}) that 
\begin{equation}
\label{eq:av0}
av_0=\left[v_n-v_m\right]_{dv/dS_w=0}\;,
\end{equation}
if the average seepage velocity has a minimum for some saturation $S_w$. 
There is, however, no {\it a priori\/} reason to believe that
this term should follow from an analytical approach based on geometry
with only two independent variables. However, the space of
extensive variables $\mathcal{M}$ is strictly speaking not complete as is. 
The statistical mechanics formalism based on Jaynes maximum entropy principle \cite{jaynesInformationTheoryStatistical1957}
developed by Hansen et al. \cite{hansenStatisticalMechanicsFramework2023,hansen2025thermodynamics}  includes configurational
entropy, and it is natural that it is included in $\mathcal{M}$. 

\subsection{A note on contact geometry}
\label{sec:contact-geometry}

As mentioned in Section \ref{sec:intro}, contact geometry \cite{arnoldMathematicalMethodsClassical1978} is the appropriate
setting for for a formalization of classical thermodynamics. The idea is to
introduce a {thermodynamic phase space} $M$ of extensive and intensive quantities in the system, which
in classical thermodynamics would be e.g. energy, entropy, volume and particle
numbers along with conjugate variables. If there are $n+1$ extensive variables, we have $n$ intensive
variables, so the total number of variables are $2n+1$. Therefore,
$\mathrm{dim}\left( M \right) = 2n+1$. All the
thermodynamic variables are initially taken to be
{independent}. One then introduces a {contact one-form}, which is
just the Gibbs one-form from thermodynamics. If we take only energy $E$, entropy
$S$ and volume $V$ as extensive variables, the contact form $\Theta$ looks like
\begin{equation}
  \label{eq:gibbs-one-form}
  \Theta \ = \ dE - \gamma_S dS + \gamma_{V}dV \;,
\end{equation}
The quantities $\gamma_S$, $\gamma_V$ are in equilibrium thermodynamics just the temperature $T$
and pressure $P$, however, they are not identified as such initially: this is only
the case on some sub-manifold $N \subset M$ that  characterizes the {equilibrium
  states} of the system. In fact, the contact form $\Theta$ defines such a
sub-manifold as $\Theta = 0$, called a {Legendre sub-manifold} \cite{arnoldMathematicalMethodsClassical1978}. More concretely, the contact form $\Theta$
 defines a
{distribution} $\mathcal{D}$ on $M$, which is just the selection of a subspace
$L_{x} \subseteq T_xM$ of the tangent space $T_xM$ at each $x \in M$. Given such a
distribution $\mathcal{D}$ stemming from the contact form $\Theta$ in equation
\eqref{eq:gibbs-one-form}, it turns out that the Legendre sub-manifolds $N \subset M$ that has
the distribution $\mathcal{D}$ as its tangent space have maximal dimension $n$. Such
sub-manifolds $N$ are more generally called called {integral sub-manifolds} \cite{spivakComprehensiveIntroductionDifferential1999,
  crampinApplicableDifferentialGeometry1987} of the distribution $\mathcal{D}$.\footnote{These sub-manifolds $N$ are also called {leaves} of the
  distribution $\mathcal{D}$.} A curve $c = c(t)$ in $M$ that lies on
$N$ can be interpreted as some quasi-static thermodynamic process. The tangent vectors to
the curve $c(t)$ are all contained in $\mathcal{D}$, which means that the curve cannot
``leave'' the equilibrium manifold $N$. It turns out that on the integral sub-manifolds $N$, we have exactly
\begin{align}
  \label{eq:gamma-equal-intensive-S}
   \gamma_{S} \vert_{N} \ =& \ \frac{\partial E}{\partial S} \;, \\
  - \gamma_{V} \vert_{N} \ =& \ \frac{\partial E}{\partial V} \label{eq:gamma-equal-intensive-S} \;,
\end{align}
in accordance with equilibrium thermodynamics. The energy $E$ is here expressed
as a function $E = E \left( S,V \right)$. In $N$, $E$ is exactly what is called
a {thermodynamic potential}.

The parallel between thermodynamics and the formalism discussed here and in 
\cite{hansenStatisticalMechanicsFramework2023,hansen2025thermodynamics} has been
developed in References \cite{hansenStatisticalMechanicsFramework2023,hansen2025thermodynamics}.
We discuss contact geometry in this context in the following, however, without including
the configurational entropy, and its conjugate, the agiture (a temperature-like variable).
We have the extensive variable $Q$ expressed as $Q = Q \left( A_w, A_n \right)$. The related contact
form is then 
\begin{equation}
  \label{eq:contact-form-Q}
 \Theta \ = \ dQ - \gamma_{A_w}dA_w - \gamma_{A_n}dA_n \;,
\end{equation}
where $\gamma_{A_w}, \gamma_{A_n}$ are identified with $\hat{v}_w$, $\hat{v}_n$
respectively on an equilibrium sub-manifold, which in our case means steady-state
flow. $v_m$ enters when the form in equation (\ref{eq:contact-form-Q}) restricted to
the steady-state manifold is rewritten as\cite{pedersenParameterizationsImmiscibleTwophase2023a} 
\begin{equation}
  \label{eq:contact-form-Q-seepage}
  \Theta \ = \ dQ - \left( v_w + S_n v_m \right)dA_w - \left( v_n - S_w v_m \right)dA_n = 0 \;.
\end{equation}
Note that formally, the quantities $S_w$ and $S_n$ must in general be treated as
independent of $A_w$, $A_n$, see Section \ref{sec:connections}. It is clear that
$v_w$ and $v_n$ in place of the thermodynamic velocities in equation
\eqref{eq:contact-form-Q} restricted to the steady state manifold would not define a Legendre
sub-manifold. $v_m$ is then a correction that brings us back to this
equilibrium sub-manifold.

In the above, we have used the assumption of extensivity of $Q$ in the remaining
extensive variables. This produces the well-known {Gibbs-Duhem relation}
\cite{hansenRelationsSeepageVelocities2018}. In a geometric context, degree-$1$
homogeneity in the extensive variables reduces the thermodynamic phase space $M$
\cite{vanderschaftLiouvilleGeometryClassical2021} in the following sense: if the
thermodynamic phase space $M$ is decomposed as $M = \mathcal{E} \times \mathcal{I} = \mathbb{R}^{n+1} \times \mathbb{R}^{n}$,
where the space $\mathcal{E}$, $\mathrm{dim}\left( \mathcal{E} \right) = n+1$, contains the
variables we denote as ``extensive'' and $\mathcal{I}$, $\mathrm{dim}\left(\mathcal{I} \right) = n$,
contains the variables we denote as ``intensive'', the homogeneity-requirement
on the extensive variables sends $\mathcal{E} = \mathbb{R}^{n+1}$ to the {quotient space} \cite{crampinApplicableDifferentialGeometry1987}  $\mathbb{P} \left( \mathcal{E}
\right) = \mathbb{P} \left( \mathbb{R}^{n+1} \right)$, the {projectivization} of $\mathcal{E}$. Thus,
projective spaces occur naturally when we introduce homogeneity, and a further study
of these types of spaces can be undertaken when working with the velocities as
introduced in Section \ref{sec:affine-space}.\footnote{Projective spaces are also
  relevant for the intensive variables: one can introduce an additional ``gauge''
variable\cite{balianHamiltonianStructureThermodynamics2001,
  arnoldMathematicalMethodsClassical1978} in $\mathcal{I}$, which is often more
convenient to work with. The intensive variables are an example of
{homogeneous coordinates} on $\mathcal{I}$, which are the standard type of
coordinates used when working with projective spaces.} 

Contact geometry is, as stated
in Section~\ref{sec:pseudo-therm-intro}, closely related to {Hamiltonian mechanics} 
\cite{arnoldMathematicalMethodsClassical1978}, which utilizes
Hamiltonian functions (which are smooth functions on phase space), which again defines
Hamiltonian vector fields. The integral curves of these vector fields yields
equations of motions for the Hamiltonian system. Similar types of relations hold
in geometric formulations of thermodynamics
\cite{arnoldMathematicalMethodsClassical1978,
  vanderschaftLiouvilleGeometryClassical2021}. In this work, a {choice of
  Hamiltonian corresponds to a choice of $Q$}. This means that the function
$Q$ itself is assumed to contain all the information about the system.

\subsection{Connections and bundle structure}
\label{sec:connections}

When introducing the description in terms of vector fields in the context of
this work, one is faced with the difficulty of making sense of expressions like
equation \eqref{eq:vm-derivative-form}. Here, the derivative operator $\partial_{S_w}$ is a
vector field. Moreover, we replace the function $v$ by a vector field $V = S_w
\partial_w + S_n \partial_{n}$. We therefore have a situation where we are evaluating the
derivative of a section $V$ in the direction of another section, $\partial_{S_w}$. This
``derivative of a section'' of the tangent bundle with respect to another
section necessitates a way of connecting the tangent spaces at different points
of the base manifold $\mathcal{M}$, since we are asking precisely how a vector field changes if
we follow it along another vector field along its integral curves on $\mathcal{M}$. Thus, we need
the general concept of a {connection} \cite{tuDifferentialGeometry2017,
  spivakComprehensiveIntroductionDifferential1999, misnerGravitation1973} on the
tangent bundle. There are many realizations of this concepts, and a thorough
treatment is outside the scope of this work. What we will say is that one way
of working with a connection is via the {covariant
derivative} \cite{misnerGravitation1973}, which measures the change in the components of
a vector field and the frame itself along another vector field. 

An important point about the
covariant derivative is that it solves the specific problem of differentiating
tangent vectors to the {tangent bundle $T\mathcal{M}$ as a whole}, and not only tangent vectors to the base space $\mathcal{M}$. To get a
tangent vector that actually lies in the tangent space, one needs a way of
``projecting'' these vectors back to the tangent space. This is often done via
the use of a {metric} \cite{misnerGravitation1973}. Note that we have not
assumed {any} type of metric structure on the space of extensive
variables or the thermodynamic phase space as a whole. This is a topic of
ongoing research (see \cite{weinholdMetricGeometryEquilibrium1975a,
  andresenThermodynamicGeometryMetrics1988} ) which is closely tied to
information theory and the Hessian of the entropy (or energy) of the system.
However, in our case we have {\it a priori\/} no knowledge of a metric, which
means that we have no idea of knowing what the contribution from such a
structure on the base space $\mathcal{M}$. We will therefore leave the discussion about
metrics here.

In the case of $V = S_w \partial_w + S_n \partial_n$ and $\partial_{S_w}$, we can form the covariant
derivative $\nabla_{\partial_{S_w}}V$, where $V$ is expressed in the coordinate frame $\left(
  \partial_w, \partial_n \right)$. Recall that we associated the general (possibly non-coordinate or
{anholonomic}) frame  $\left(
  \mathbf{e}_w, \mathbf{e}_n \right)$ to the seepage velocities. $V$ expressed in this frame is
then just $V = S_w \mathbf{e}_w + S_n \mathbf{e}_n$. A general expression for
the covariant derivative using an arbitrary frame $\left\{ \mathbf{e}_i \right\}$
and vector fields $X$, $u$
is \cite{misnerGravitation1973}
\begin{align}
  \label{eq:covariant-derivative}
  \nabla_{u} X \ =&\left(u^{j}\mathbf{e}_i\nabla_{\mathbf{e}_j} \left( X^{i} \right) + u^{j}X^i \nabla_{\mathbf{e}_j} \left( \mathbf{e}_i \right) \right) \nonumber \\
  =& \left(u^j \mathbf{e}_j\left( X^k \right) +  X^iu^j \Gamma^k_{ji} \right)\mathbf{e}_{k} \ \;,
\end{align}
\noindent where $\Gamma^k_{ji}$ are the {connection
  coefficients}\footnote{These are, given some additional assumptions, just what we call the
  {Christoffel symbols}\cite{misnerGravitation1973}.}  of the
connection with respect to the basis, and the notation $\mathbf{e}_j \left( v^k \right)$
(for all indices) denotes the action of the frame element $\mathbf{e}_j$ on the function
$v^{k}$. If we first set $V = S_w \partial_w + S_n\partial_n$ so that $\left\{ \mathbf{e}_i \right\} = \left( \partial_{S_w},
  \partial_{S_n} \right)$, $X^i = V^{i} = \left( S_w, S_n \right)$, and $u = \partial_{S_w} = \left(
  A_w + A_n \right)\left(
  \partial_{w} - \partial_{n} \right)$ so that $u^{i} = \left( A_w + A_n, - A_w + A_n
\right)$, one can show that the covariant derivative $\nabla_{\partial_{S_w}} V$ reduces to
$\left( \partial_w - \partial_n \right)$, which applied to $Q$ yields $\hat{v}_w - \hat{v}_n$.
However, if if we change the frame from the coordinate frame to the
seepage-frame, $\left( \partial_w, \partial_n \right) \mapsto \left(
  \mathbf{e}_w, \mathbf{e}_n \right)$, the last term in both lines of equation
\eqref{eq:covariant-derivative} are not necessarily zero. In fact, the first
term of the first line of equation \eqref{eq:covariant-derivative} can be written as
$v_w - v_n$, and the second term is exactly equation
\eqref{eq:co-moving-convex-comb}. In general, if the frame is a
non-coordinate-frame, the connection coefficients $\Gamma^{k}_{ji}$ contains
contributions from both the metric and the {commutation coefficients} \cite{misnerGravitation1973}  of
the frame, which describes exactly the dependency of the frame
elements. Thus, {we can connect the co-moving velocity to the existence of a type
of metric, the dependency between the frame elements, or both}.  \\

We can draw an analogy between equation \eqref{eq:co-moving-convex-comb} and the
connection-term $u^jX^i\nabla_{\mathbf{e}_j} \left( \mathbf{e}_{i} \right)$ in equation
\eqref{eq:covariant-derivative}. This term contains the derivatives of the frame
elements with respect to $\partial_{S_w}$, which are expanded in the frame itself to
yield the connection coefficients. If we instead stick to the first line in equation~\eqref{eq:covariant-derivative}, using the relation $\mathbf{e}_w \sim v_w$,
$\mathbf{e}_n \sim v_n$, we see that a term $\partial_{S_w} v_{\alpha}$ in equation \eqref{eq:co-moving-convex-comb} is analogous to the
covariant derivative of a single vector in the frame $\left\{ \mathbf{e}_{\alpha} \right\}$, so we have $\partial_{S_w} v_{\alpha}
\sim \nabla_{\partial_{S_w}}\mathbf{e}_{\alpha}$. Thus, the vector field $V_m$ associated to $v_m$
can be written as
\begin{equation}
  \label{eq:Vm-covariant-form}
  V_m \ = \ S_w \nabla_{\partial_{S_w}} \mathbf{e}_w + S_n \nabla_{\partial_{S_w}} \mathbf{e}_n \;.
\end{equation}
This is a well-defined vector field since $\nabla_{\partial_{S_w}} \mathbf{e}_\alpha$ produces
vector fields, and a linear combination of vector fields is again a vector
field. Moreover, since the functions $S_w, S_n$ on $\mathcal{M}$ satisfy $S_w + S_n = 1 $,
equation \eqref{eq:Vm-covariant-form} is defined on an affine bundle. An affine bundle has no
preferred zero-section, and the only expressions that are independent of the
choice of zero-section are affine combinations of sections. The situation is
therefore analogous to affine combinations which are independent of the choice of
origin \cite{rockafellarConvexAnalysis1997}. The vector fields related to both equation \eqref{eq:v-both-definitions}
and equation \eqref{eq:Vm-covariant-form} share the property that they are
affine combinations of sections, and are therefore independent of any choice of origin
in the spaces of velocities. Since they are independent of the zero-section, we
could use these vector fields themselves as zero-sections. The defining
difference between vector bundles and affine bundles are that vector bundles
always has a zero-section, so defining a zero-section of the affine bundle is
equivalent to a vector bundle.

The relation between these considerations, \cref{eq:vw-minu-vn-simpler} and the
discussion after \cref{eq:vm-v-prime-affine} can
be formulated in terms of the connection. We will only provide an explanation on
a conceptual level, as a thorough treatment is outside the scope of this work.

On a one-dimensional manifold, the only possible form of the covariant
derivative is $g \left(x \right) \partial_x$. \footnote{The covariant derivative can be
  seen as a projection operator that projects the tangent vectors to the tangent
  bundle $T\mathcal{M}$ itself onto the tangent spaces of $\mathcal{M}$.}. We have here a single
variable $S_w$, which in reality is a parameter along a line embedded in a
higher dimensional space. Let this space be two-dimensional as before, with the
same frame elements $\left\{ \mathbf{e}_i \right\}$, $i = w,n$ as we have
already considered. View $V_m$ as the zero-section of an affine bundle. On the
level of bundles, an arbitrary zero-section is handled by a solder form
\cite{saundersGeometryJetBundles1989} \footnote{ In particular, a connection on
  an affine bundle is an example of an affine connection, of which the covariant
  derivative is one manifestation. The connection on the affine bundle is in
  this case an example of a more general definition of a connection called a
  Cartan connection \cite{iveyCartanBeginnersDifferential,
    ehlersNonholonomicSystemsMoving2007}.}.

On the tangent bundle, a solder form represents a relation between the tangent
space at a point and the vertical spaces of the bundle $T\mathcal{M}$. The vertical space
at a point $p$ of $T\mathcal{M}$ consist of all tangent vectors to $T\mathcal{M}$ that project to
tangent vectors of $\mathcal{M}$. These spaces form a bundle called the vertical bundle
$VT\mathcal{M}$. A solder form $\tau$ at a point $x \in \mathcal{M}$ is defined in terms of a distinguished
section, here $V_m$, and is a linear map (not affine)
\begin{equation}
  \label{eq:solder-form}
  \tau_x \;: T_x \mathcal{M} \rightarrow V_{V_M \left( x \right)}T\mathcal{M} \;.
\end{equation}
Intuitively, $\tau_x$ can be seen as relating the tangent vectors at $T_x\mathcal{M}$ with
all tangent vectors of the entire bundle $T\mathcal{M}$ that project to vectors on $T_x\mathcal{M}$.
This is one way of formulating the necessary freedom in the transformation
between thermodynamic- and seepage velocities, expressed in terms of bundles.

Due to the linearity of the map $\tau_x$, the solder form can be incorporated into
the covariant derivative, where its effect enters into the connection
coefficients $\Gamma^k_{ij}$ in \cref{eq:covariant-derivative}. The connection
coefficients contain contributions from a metric (which can be zero), in
addition to terms that can arise if the frame is anholonomic, i.e. not a
coordinate frame. This is the case for a solder form, which enter into the
connection coefficients as this latter type of term. These are the terms that
give rise to torsion of the connection. In terms of the frame $\left\{ \mathbf{e}_i
\right\}$, an often used picture of torsion of a connection is to parallel
transport the frame vectors along each other some unit distance. If the two
parallel-transported vectors and the two frame vectors form a closed
parallelogram, the connection is free of torsion \cite{misnerGravitation1973}
\footnote{This is a bit imprecise, as how fast the gap between the transported
  vectors matter as the distance they are transported increases. It poses no
  harm to ignore this here.}.

The takeaway is that in our
case, \cref{eq:covariant-derivative,eq:Vm-covariant-form} are not mutually
exclusive, as $V_M$ can be included into \cref{eq:covariant-derivative}. We then
have several ways of viewing $V_M$: either as related to a solder form, or more
generally the connection coefficients of some frame, or as induced by some
metric. In conclusion, a differential geometric treatment that allows for additional
``translational'' terms in \cref{eq:vm-v-prime-affine,eq:vw-minu-vn-simpler}
is much more involved, and depends on how the frame $\left\{ \mathbf{e}_i
\right\}$ is defined.

\section{Conclusion}
\label{sec:conclusion}

We have in this work introduced basic geometric ideas into the analysis of a
pseudo-thermodynamic description of two-phase flow in porous media. The goal was
to pave the way for the usage of geometry in interpreting and classifying
relations occurring in the theory and in equilibrium thermodynamics in general. A
relatively terse introduction of necessary concepts was presented in the context
of our choice of extensive and intensive variables, in addition to the
underlying assumption of degree-$1$ Euler homogeneity of the total volumetric
flow rate in the extensive variables. We have in this endeavour provided two
potential routes of further study of the relations presented in
Section~\ref{sec:rea}. One is to apply the language of classical affine and
projective geometry and work directly with functional values of the velocities.
To the authors' knowledge, this approach is new in the context of two-phase flow
in porous media, and uncommon in the study of thermodynamics in general.\footnote{One only
  implicitly uses this structure when considering {specific} quantities
  in thermodynamics.}

The second route is to bring the formalism closer to contemporary formulations
of the geometric structure of thermodynamics. The approach in this article was a
natural continuation of investigating vector spaces and coordinates on the space
of extensive variables in previous work
\cite{pedersenParameterizationsImmiscibleTwophase2023a}.

On the subject of continued work on tangent vector fields as presented in this
work, it would be interesting to see how the terms $\sim a$ in equation
\eqref{eq:vm-constitutive} can appear if more variables are included in the
space of extensive variables. Even though the framework presented here cannot claim any
predictive power for the parameters $b$,$a$ in the constitutive equation for
$v_m$, it aids in gaining intuition for what the co-moving velocity and other
relations in Section \ref{sec:rea} represents geometrically. Future work on the
theoretical basis of the pseudo-thermodynamic theory of two-phase flow which
applies more standard formalism used in geometric equilibrium thermodynamics,
(for instance contact geometry), the geometric concepts introduced here still
applies. Moreover, a separate study in terms of contact geometry is highly relevant.

In the classical description where points of the affine space of velocities were
identified with numbers, the most natural way forward is to formulate the theory
in terms of projective geometry. Projective geometry is a particularly rich and well-known
topic in both mathematics and physics,
and presents many avenues of exploration. One of these could be to try to
explicitly compute an invariant of projective geometry, the so-called
{cross-ratio} \cite{bergerGeometry1994a,
  richter-gebertPerspectivesProjectiveGeometry2011}, from the values of the
velocities. One could use this to investigate the assumption of homogeneity in more
detail, and possibly use projective relations as a guide to obtain new constitutive
relations for the co-moving velocity.\\

\textbf{Author Contributions}:  Conceptualization, H.P.; Validation, A.H.; Writing—original draft, H.P.; Writing and editing, A.H. All authors have read and agreed to the published version of the manuscript. \\

\textbf{Funding}: This work was partly supported by the Research Council of Norway through its 
Centers of Excellence funding scheme, project number 262644.  AH furthermore 
acknowledges funding from the European Research Council (ERC) through grant 
agreement 101141323 AGIPORE. \\

\textbf{Data Availability Statement: }The original contributions presented in this study are included in the article. Further inquiries can be directed to the corresponding author. \\

\textbf{Acknowledgments:} We thank Santanu Sinha for valuable discussions and comments. \\

\textbf{Conflicts of Interest:} The authors declare no conflicts of interest.
\bibliography{affine}


\end{document}